# Structure-related bandgap of hybrid lead halide perovskites and close-packed $APbX_3$ family of phases

Ekaterina I. Marchenko [a,c], Sergey A. Fateev [a], Vadim V.Korolev [b], Vladimir Buchinskii [c], Eremin N.N.[c], Eugene A. Goodilin[a,b] and Alexey B. Tarasov*[a,b]

[a] *Laboratory of New Materials for Solar Energetics, Department of Materials Science, Lomonosov Moscow State University; 1 Lenin Hills, 119991, Moscow, Russia;*
*e-mail:alexey.bor.tarasov@yandex.ru*
[b] *Department of Chemistry, Lomonosov Moscow State University; 1 Lenin Hills, 119991, Moscow, Russia*
[c] *Department of Geology, Lomonosov Moscow State University; 1 Lenin Hills, 119991, Moscow, Russia*
[*] corresponding author

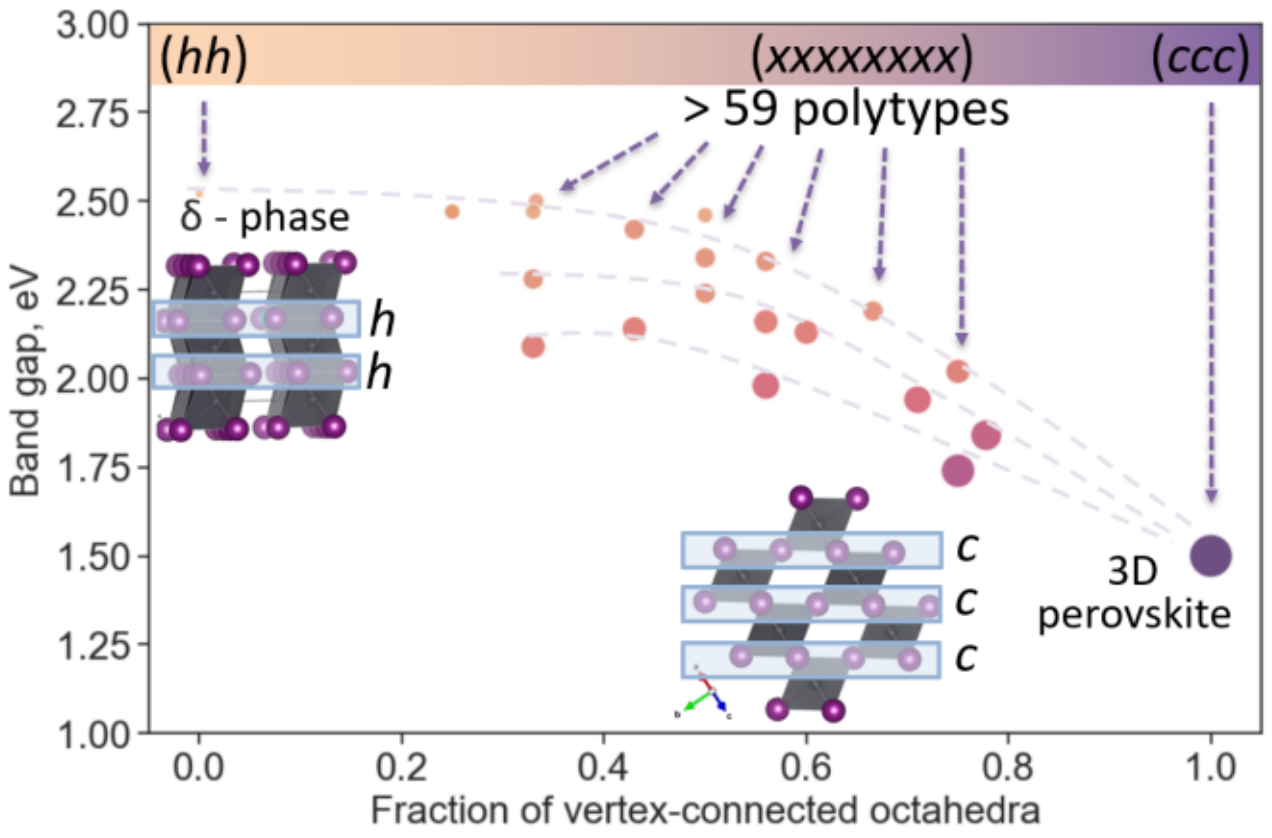

**Abstract**

Metal halide perovskites $APbX_3$ ($A^+$ = $FA^+$ (formamidinium), $MA^+$ (methylammonium) or $Cs^+$, $X^-$ = $I^-$, $Br^-$) are considered as prominent innovative components in nowadays perovskite solar cells. Crystallization of these materials is often complicated by the formation of various phases with the same stoichiometry but structural types deviating from perovskites such as well-known the hexagonal delta $FAPbI_3$ polytype. Such phases are rarely placed in the focus of device engineering due to their unattractive optoelectronic properties while they are, indeed, highly important because they influence on the optoelectronic properties and efficiency of final devices. However, the total number of such phases has not been yet discovered and the complete configurational space of the polytypes and their band structures have not been studied systematically. In this work, we predicted and described all possible hexagonal polytypes of hybrid lead halides with the $APbI_3$ composition using the group theory approach, also we analyzed theoretically the relationship between the configuration of close-packed layers in polytypes and their band gap using DFT calculations. Two main factors affecting the bandgap were found including the ratio of cubic (*c*) and hexagonal (*h*) close-packed layers and the thickness of blocks of cubic layers in the structures. We also show that the dependence of the band gap on the ratio of cubic (*c*) and hexagonal (*h*) layers in these structures are non-linear. We believe that the presence of such polytypes in the perovskite matrix might be a reason for a decrease in the charge carrier mobility and therefore it would be an obstacle for efficient charge transport causing negative consequences for the efficiency of solar cell devices.

## Introduction

Metal halide perovskites with a general formula $ABX_3$, where $A^+$ is $Cs^+$, $CH_3NH_3^+$ ($MA^+$), $CH(NH_2)_2^+$ ($FA^+$), $B^{2+}$ is $Pb^{2+}$ or $Sn^{2+}$, $X^-$ is $I^-$, $Br^-$, have attracted significant attention as materials for photovoltaic and optoelectronic applications [1]. In contrast to traditional semiconductors, halide perovskites are featured by the ease of their synthesis with superior optoelectronic properties using simple low-temperature wet chemistry and high tunability of their physical properties. Due to the “soft” ionic framework, these materials exhibit a significant polymorphism, which manifests as an existence of several phases with different structures within the same $ABX_3$ composition. For example, phase transitions between high-temperature low-bandgap (1.5 eV) photoactive cubic perovskite and the low-temperature high-bandgap (>2.8 eV) hexagonal phase are well known for widely used $FAPbI_3$ and $CsPbI_3$ compositions [2], [3], moreover, all polymorphs are featured by quite similar formation energies [4]. Such a striking variety in the optoelectronic properties of polymorphs significantly complicates the achievement of desired characteristics of perovskite thin films due to competitive crystallization of different phases from solution. Moreover, according to recent studies [2], [3], the crystallization of halide perovskite films can proceeds through a number of intermediates with the structure of $APbX_3$ polymorphs [2], [3]. The most common of them is hexagonal 2H polytype typical for the compounds with relatively “large” organic cations [5] known as δ-$FAPbI_3$ in the case of formamidinium

– the yellow phase of $FAPbI_3$ existing below 185 °C [6], [7]. Gratia et al. reported that phase evolution of $(FAPbI_3)_{1-x}(MAPbBr_3)_x$ perovskite polycrystalline films proceeds through the sequence of polymorphs from 2H to 4H, 6H and finally to 3C-$(FAPbI_3)_{1-x}(MAPbBr_3)_x$ (3C - cubic phase) [2]. Recently, a new intermediate of 8H phase was discovered, showing even a more complex character of phase evolution of FA-containing hybrid halide perovskites [3]. As far as we know, there are 5 polytypes of $APbX_3$ ($X=I^-$, $Br^-$) in total observed experimentally: 2H, 3C, 4H, 6H and 8H (Figure 1). All these compounds are characterized by close packing of the monovalent cations $A^+$ together with the halogen anions $X^-$, where $Pb^{2+}$ cations occupy ¼ of the octahedral voids and form $[PbX_6]$ octahedra connected into a framework (Figure 1). However, the complete configurational space of possible polytypes and their band structures have not been systematically studied.

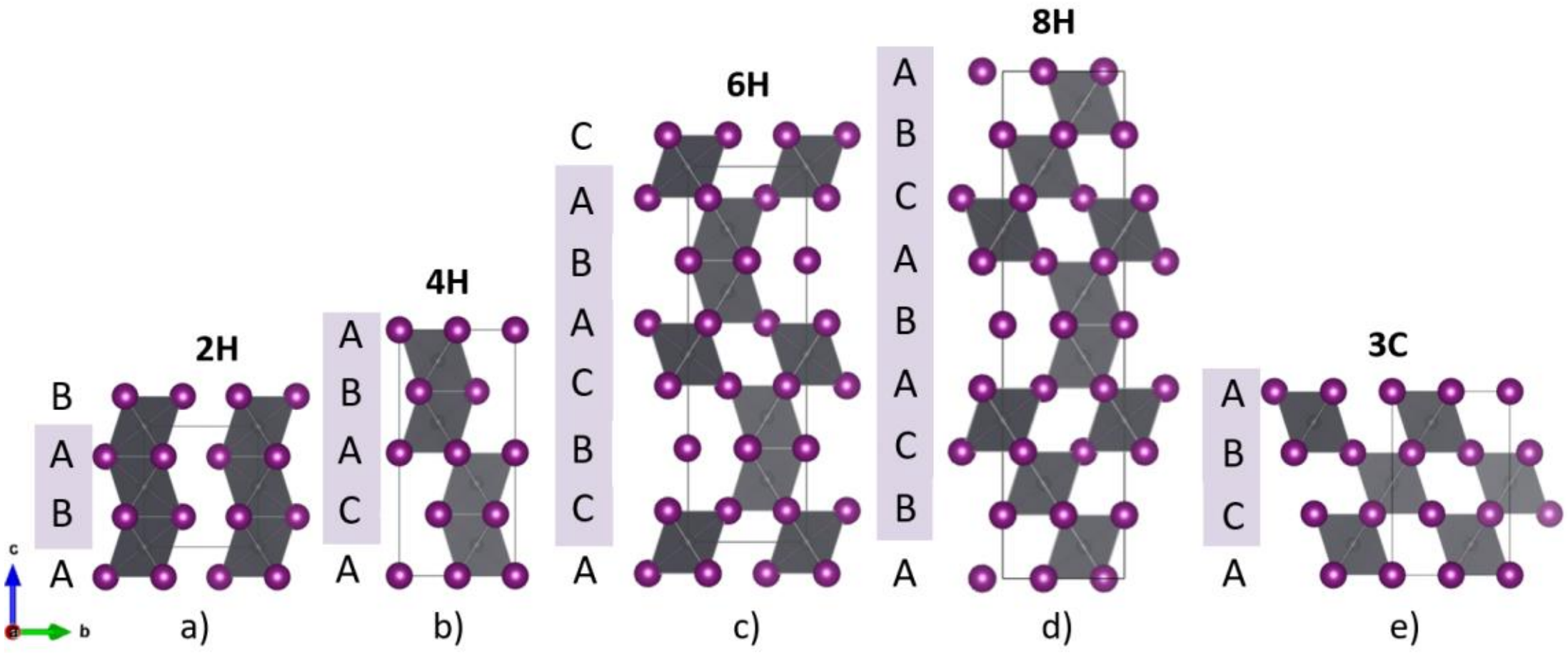

Figure 1. The topology of experimentally known $APbX_3$ close-packed polymorphs accordingly [2] and [3]: (a) 2H, (b) 3C – cubic structure is shown in hexagonal axes, (c) 4H, (d) 6H, (e) 8H.

We define the aforementioned polymorphs as structures intermediate between the 2H (hexagonal) and 3C (cubic) phases with the $ABX_3$ composition containing a different number of cubic close-packed layers with corner-sharing $PbX_6$ octahedra and hexagonal ones with face-sharing octahedra. Obviously, different types of octahedra connectivity should significantly affect the optoelectronic properties of these compounds. Although the optical properties of the known FA-based polymorphs were not measured except for 3C and 2H phases, the different color of the compounds [2], [3] unambiguously implies the different bandgaps. In particular, the successive color change from yellow for 2H to orange for 4H, and then red and black for 6H and 8H, respectively, indicates a gradual decrease in the bandgap in this series. A more complex trend is observed for tin iodide polytypes with different organic cations as found by Stoumpos [5]. While 2H, 4H, and 6H are pale yellow, orange, and dark red, respectively, the color of 9R polytypes varies from yellow to orange. Thus, a relatively small set of experimentally studied polytypes misleads the understanding of relationships between crystal structures and optical properties for this class of materials.

It is important to note that polymorphs can also have a significant effect on the functional properties of the devices based on 3D perovskites thin films. For example, in polycrystalline films of FA-based perovskites, polymorphs can form a bulk phase impurity [2] or exist in a "hidden form" of stacking faults manifested as an "admixture" of hexagonal close-packed layers [8]. Another possible evidence of "hidden" polymorphs are the facts of a gradual shift of a photoluminescence (PL) peak along the interface between the 2H and 3C phases of $FAPbI_3$ measured *in-situ* during phase transition in single crystal (1) [9] and the shift of PL for polycrystalline film of the 2H phase annealed under different conditions (2) [10].

For sure, a set of $ABX_3$ polytypes, which are intermediate phases between 2H and 3C, is not limited to the suite of experimentally discovered structures. The most of them are difficult to study experimentally, since it is difficult to obtain these compounds in a form of single-phase samples. A number of polytypes with low ordering energies (around thermal energy at room temperature), which could form distinct phases or appear as stacking faults, were predicted theoretically [4], however, the real number of $CsPbX_3$ polytypes with different close packed layers of packages has never been univocally determined. Obviously, there is no clear understanding of the relationship between crystal and electronic structures in the entire possible crystallographic configurations of such polytypes.

In present work, we assessed a full configurational space of $APbX_3$ polytypes with close-packed structures from a crystallographic point of view using a group theory approach and the analyze of the dependence of band gaps on the type of close packings in the polytypes based on DFT calculations.

## Methods

### ***Assessment of possible close-packed $APbI_3$ crystal structures***

Close-packed layers of ions in close-packed structures are presented by two types of motifs: hexagonal (*h*) and cubic (*c*). In the *h* layers, all ions are located on the mirror plane of symmetry (*m*) between adjacent layers of the close-packing while all ions are placed in inversion centers ($\overline{1}$) in the *c* type layers. To construct the whole possible configurations of stoichiometric $APbX_3$ polytypes with the given number of close pack layers (n), we used the following algorithm:

1) Selection of all n/3 combinations of the *h* and *c* symbols with the even number of h and $\sum c$ not divisible by 3. This results (after tripling) in all rhombohedral packings of the period n with the symmetry of $C_{3v}^5$=$R3m$, $D_{3d}^5$=$R\overline{3}m$ ($O_h^5$ = $Fm\overline{3}m$).
2) Consideration of all *n* combinations of *h* and *c* symbols with the odd number of *h*. As a result, we get all packings with $6_3$ axes with $C_{6v}^4$=$P6_3mc$ and $D_{6h}^4$=$P6_3/mmc$ symmetry.
3) Choice of all *n* combinations of *c* and *h* symbols with an even number of *h* and $\sum c$ divisible by 3. As a result, we obtain packings with $C_{3v}^1$=$P3m$, $D_{3d}^3$=$P\overline{3}m$ and $D_{3h}^1$=$P\overline{6}m2$ symmetry.
4) Construction of the obtained close packings from iodine and cesium ions and fill ¼ octahedral voids by lead ions.

All possible periodic close packings with the number of layers (n) from 2 to 11 and their characteristics are shown in Table 1 and Table S1. Interestingly, the close packing with the minimal possible symmetry group for undistorted close packings *P*3*m* is found for the first time among nine-layer packings (9H(5), 9H(6), 9H(7)) (Figure S1). It is also worth noting that the close packing with the $P6_3mc$ symmetry is observed also for the first time among 12-layer packings, and the close packing with the *R*3*m* symmetry is newly found only among 21-layer packages. These space groups of close packings are predominant with multilayer ones.

There is the only one configuration of 2H, 3C, 4H and 5H close packings, the number of possible configurations grows rapidly further becoming 2 for 6H, 3 - 7H, 6 - 8H, 7 - 9H, 16 - 10, 21 - 11H. Further, the number of possible packages dramatically increases, so, for 12H there are 43 variants.

We construct the crystallographic information files of $APbI_3$ polytypes of close-packed structures with *n* from 2 to 11 (see the SI files). The packing symbol is written in several notifications: *hc* and *ABC* letters (Table 1). The $PbI_6$ octahedra in these structures can only be connected along vertices and faces while maintaining the $APbX_3$ composition. The unit cell parameters *a* and *b* of the hexagonal polytypes were 8.8 Å, which corresponds to crystal structures with formamidinium cations. Visualization of the structures was carried out using the VESTA program [11]. All crystal structures of the constructed polytypes, simulated diffraction patterns and their band structure diagrams are given in Supporting Information files (Figures S1-S14).

### ***Band structure calculations***

The electronic band structures were calculated by means of the density functional theory (DFT) formalism [12], [13] implemented in the GPAW code [14], [15]. The standard projected-augmented wave [16] (PAW) datasets were used to describe the core-valence interaction. The electronic wave functions were expanded in a plane-wave basis set up to an energy cut-off of 500 eV. The Perdew–Burke–Ernzerhof [17] (PBE) exchange-correlation functional was used including spin-orbit (SOC) coupling unless otherwise stated. Initially, the Brillouin zone sampling was done on a Γ-centered grid with a density of 8 *k*-points per $Å^{-1}$ according to the Monkhorst–Pack [18] scheme; the Fermi–Dirac smearing of 0.05 eV was employed. Then, the Brillouin zone along high symmetry directions was sampled with a density of 50 *k*-points per $Å^{-1}$ as suggested by Setyawan and Curtarolo [19]. The calculations were organized using the Atomic Simulation Recipes framework [20].

It is well known, that strong SOC effects in hybrid perovskites significantly lower the band gap [21]. In Figure S2 a-b we compare the results of band structure calculations for one of the 9H polytypes with and without SOC. It is clearly seen that the value of the band gap is most consistent with the results of calculations without taking into account SOC, while the shape of the bands and their dispersion are better reproduced as a result of calculations of the band diagram with SOC, which was also shown previously in a number of works [21], [22]. The band gap energy in hybrid perovskites is almost fully determined by the configuration of the inorganic sublattice [23]–[25] whereas organic moieties affect the band gap indirectly, only through the distortion of inorganic layers. In this case, the following approach works: formamidinium cations can be replaced by Cs atoms to calculate the band gap [26], [27], [28].

## Results and discussion

To reveal structure-properties relationships in $APbI_3$ polytypes ($A^+$ – $Cs^+$, $FA^+$) we constructed the crystal structures with all crystallographically possible periodic close packings with the number of close packed layers (n) from 2 to 11 (Table 1 and Table S1). There are only 8 space groups of undistorted close packings: $P3m$, $P\bar{6}m2$, $P6_3mc$, $P\bar{3}m$, $R3m$, $R\bar{3}m$ ($Fm\bar{3}m$), $P6_3/mmc$. The symmetry of undistorted close-packings of $APbI_3$ polytypes interrelated with each other by the group-subgroup relationship shown in Figure 2. The minimal possible space group of undistorted multilayer packing corresponds to the space group $P3m$. If one doubling symmetry element ($m_z$, $2_1$ or $\bar{1}$) is added to the minimal subgroup ($P3m$), the close-packing symmetry changes to $P\bar{6}m2$, $P6_3mc$, $P\bar{3}m$ respectively. If doubling symmetry elements are added to the $P\bar{6}m2$, $P6_3mc$, $P\bar{3}m$ space groups, the close packing space groups increase to the maximal hexagonal supergroup of undistorted close packings − $P6_3/mmc$. An addition of the translation vectors of the *R* lattice to the minimal subgroup $P3m$ raises the space group to *R*3m or it becomes $R\bar{3}m$ together with a mirror plane $m_z$. The distribution of $ABX_3$ close-packed polytypes by symmetry of close-packages is shown in Figure 2. Interestingly, the $P6_3mc$ symmetry is found for the first time only among 12H close packages, and $R3m$ appeared for the 21H packing only.

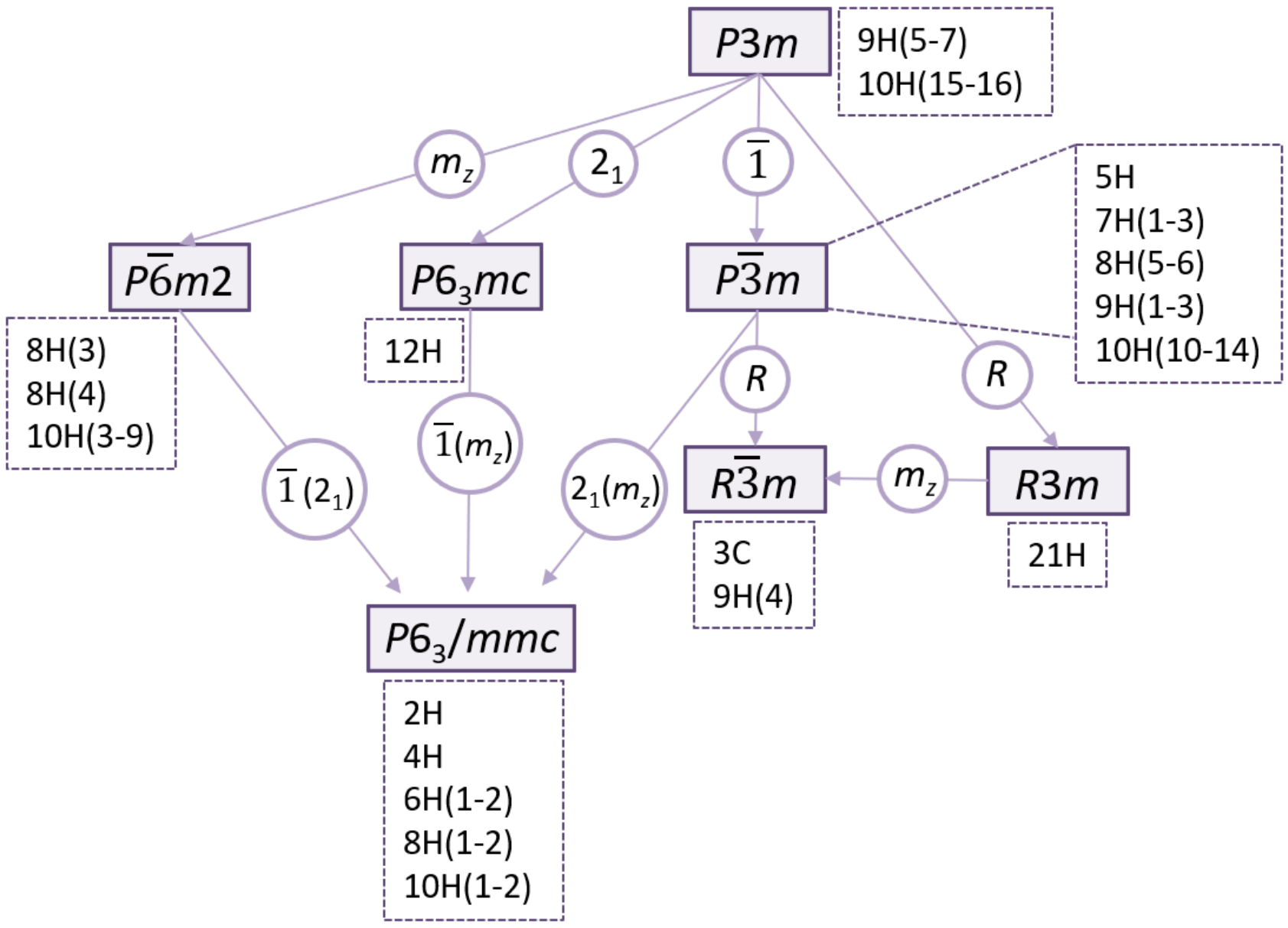

Figure 2. Symmetry relationships between undistorted close packings of $ABX_3$. The dotted line shows the symmetry of polytype packings with the number of layers (n) from 2 to 10. Space group $P6_3mc$ occurs for the first time among 12H and $R3m$ among 21H polytypes. The numbers in parentheses indicate the serial numbers of the polytypes.

Table 1. Characteristics of possible configurations of close packings from 2 to 9 layers for $APbI_3$ in different notifications.

| A number of layers in close packing | (ABC) notification | (*hc*) notification | Space group symmetry of close-packings | Fraction of closed-packed layers where $PbI_6$ octahedra connected by vertices | Calculated band gap (without SOC), eV | Calculated band gap (direct) (without SOC), eV | Known experimentally |
|---|---|---|---|---|---|---|---|
| 2H | (AB) | \|hh\| | $D_{6h}^4$ ($P6_3/mmc$) | 0 | 2.52 | 2.95 | yes [2] |
| 3C | (ABC) | \|ccc\| | $D_{3d}^5$ ($R\bar{3}m$) | 1 | 1.50 | 1.50 | yes [2][3] |
| 4H | (ABAC) | \|chch\| | $D_{6h}^4$ ($P6_3/mmc$) | 0.5 | 2.46 | 2.51 | yes [2][3] |
| 5H | (ABCAB) | \|hccch\| | $D_{3d}^3$ ($P\bar{3}m$) | 0.6 | 2.13 | 2.15 | no |
| 6H(1) | (ABACBC) | \|cchcch\| | $D_{6h}^4$ ($P6_3/mmc$) | 0.6666 | 2.19 | 2.19 | yes [3][2] |
| 6H(2) | (ABABAC) | \|chhhch\| | $D_{6h}^4$ ($P6_3/mmc$) | 0.3333 | 2.50 | 2.64 | no |
| 7H(1) | (ABABACB) | \|hhccchh\| | $D_{3d}^3$ ($P\bar{3}m$) | 0.43 | 2.14 | 2.14 | no |
| 7H(2) | (ABCBACB) | \|hchcccc\| | $D_{3d}^3$ ($P\bar{3}m$) | 0.71 | 1.94 | 1.94 | no |
| 7H(3) | (ABCBCAC) | \|cchhchh\| | $D_{3d}^3$ ($P\bar{3}m$) | 0.43 | 2.42 | 2.42 | no |
| 8H(1) | (ABCABACB) | \|hccchccc\| | $D_{6h}^4$ ($P6_3/mmc$) | 0.75 | 2.02 | 2.02 | yes[3] |
| 8H(2) | (ABABACAC) | \|chhhchhh\| | $D_{6h}^4$ ($P6_3/mmc$) | 0.25 | 2.47 | 2.70 | no |
| 8H(3) | (ABABABCB) | \|hhhhhchc\| | $D_{3h}^1$($P\bar{6}m2$) | 0.25 | 2.47 | 2.62 | no |
| 8H(4) | (ABABCACB) | \|hhhcchcc\| | $D_{3h}^1$($P\bar{6}m2$) | 0.5 | 2.24 | 2.24 | no |
| 8H(5) | (ABCBCACB) | \|hchhchcc\| | $D_{3d}^3$ ($P\bar{3}m$) | 0.5 | 2.34 | 2.37 | no |
| 8H(6) | (ABACBACB) | \|hhcccccc\| | $D_{3d}^3$ ($P\bar{3}m$) | 0.75 | 1.74 | 1.74 | no |
| 9H(1) | (ABCABCBAC) | \|cccchcch\| | $D_{3d}^3$ ($P\bar{3}m$) | 0.78 | 1.84 | 1.84 | no |
| 9H(2) | (ABCBACACB) | \|hchcchhcc\| | $D_{3d}^3$ ($P\bar{3}m$) | 0.56 | 2.33 | 2.34 | no |
| 9H(3) | (ABABABACB) | \|hhhhhhccc\| | $D_{3d}^3$ ($P\bar{3}m$) | 0.33 | 2.09 | 2.15 | no |
| 9H(4) | (ABABCBCAC) | \|chhchhchh\| | $D_{3d}^5$ ($R\bar{3}m$) | 0.33 | 2.47 | 2.62 | no |
| 9H(5) | (ABCABACAC) | \|ccccchchhh\| | $C_{3v}^1$ ($P3m$) | 0.56 | 1.98 | 1.98 | no |
| 9H(6) | (ABCACABAC) | \|cchhchch\| | $C_{3v}^1$ ($P3m$) | 0.56 | 2.16 | 2.16 | no |
| 9H(7) | (ABABACACB) | \|hhhhchhcc\| | $C_{3v}^1$ ($P3m$) | 0.33 | 2.28 | 2.28 | no |

First of all, there are much more possible $APbI_3$ closed-packed polytypes than reported so far in experimental studies [28][2]. For example, 4H, 6H(1) and 8H(1) experimentally observed polytypes contain two h close-packed layers between c (cubic layers) per unit cell and are not the exhaustive number of close-packed polytypes with two hexagonal layers between the 3C perovskite and the 2H delta polytype. Other polytypes with two hexagonal layers between cubic layers per unit cell, that have not yet been experimentally discovered, would be added to this series, such as 7H(2), 9H(1), 10H(1), 11H(1), 12R and so on. For all the constructed crystal structures with the number of layers from 2 to 11, we provide their unit cell parameters, atomic coordinates and theoretical diffraction patterns given in Supporting information.

The considered polytypes with different and the same number of closed-packed layers differ in the ratio of *c* and *h* layers (see Tables 1 and S1) representing the corner-shared and face-shared types of octahedral connectivity, respectively. A different type of connection leads to the difference in the symmetry of atomic orbitals overlapping and, therefore, govern the electronic structure of polytypes through the ratio of cubic and hexagonal layers. For example, a fraction of cubic layers in two possible 6H polytypes is 2/3 and 1/3 for 6H(1) and 6H(2) polytypes respectively, which leads to a difference in the calculated band gap by 0.31 eV (2.19 eV for 6H(1) and 2.5 eV for 6H(2)). With an increase in the number of layers *n* of the close packings in the possible polytypes, the variety of band gap values among polytypes with the same *n* will increase (Figure 3). Thus, for 8H and 9H polytypes, the difference between the maximum and minimum values of the band gap is about 0.7 eV.

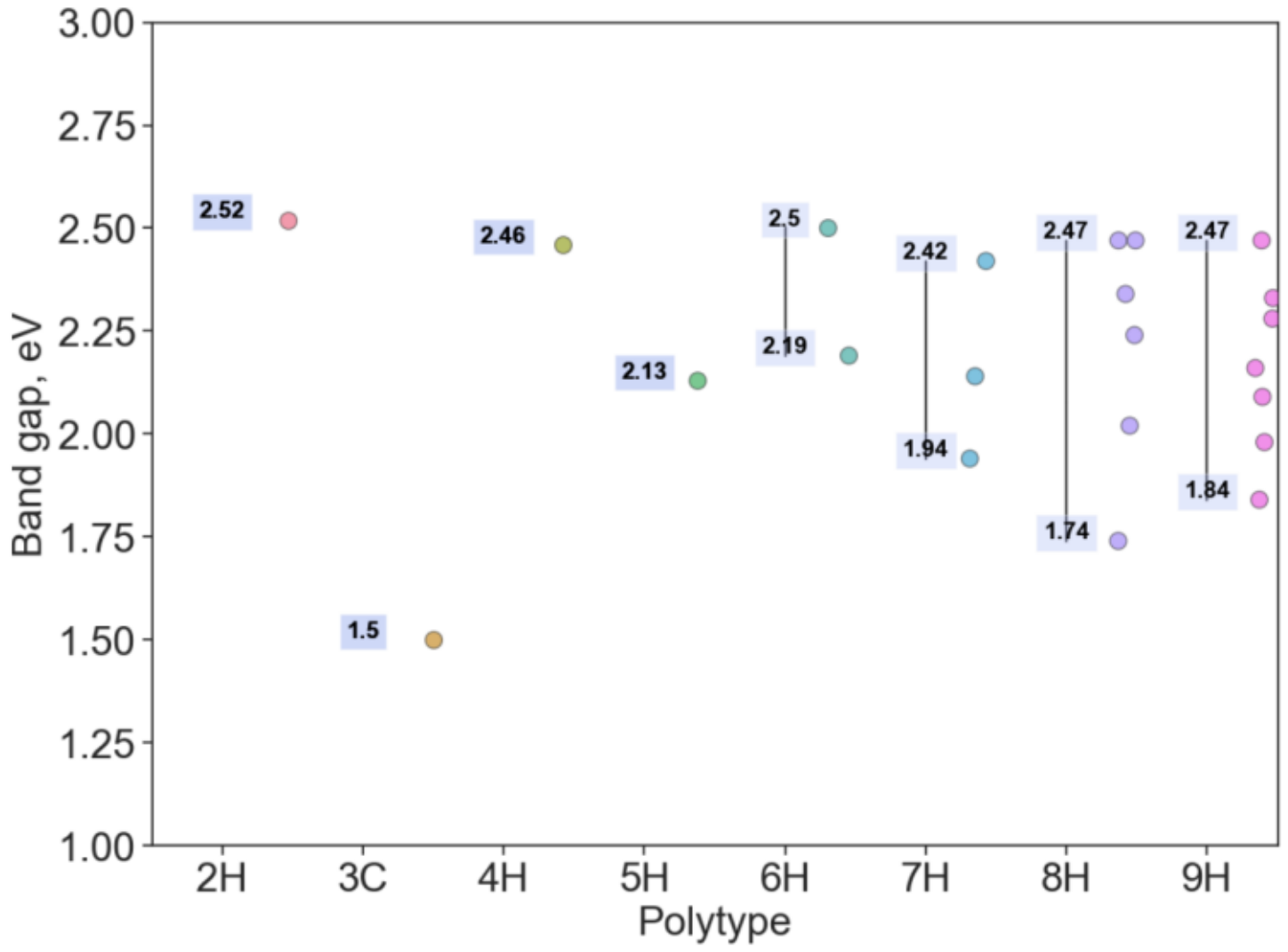

Figure 3. The strip plots of the calculated band gap from the number of close-packed layers in polytypes. Color dots show the value of the calculated band gap without SOC for a particular polytype. The numbers show the maximum and minimum calculated values of the band gap for the polytypes with the same close-packed.

The band gap in the studied multilayer polytypes with different close packings varies from 2.52 eV (for 2H polytype), where all octahedra are connected along faces to 1.5 eV (for 3C polytype) if all the octahedra are connected by vertices. As clearly seen from the Figure 4, the dependence of the number of *h* and *c* layers in the structure is not linearly dependent on the number of the close-packed layers in the polytype.

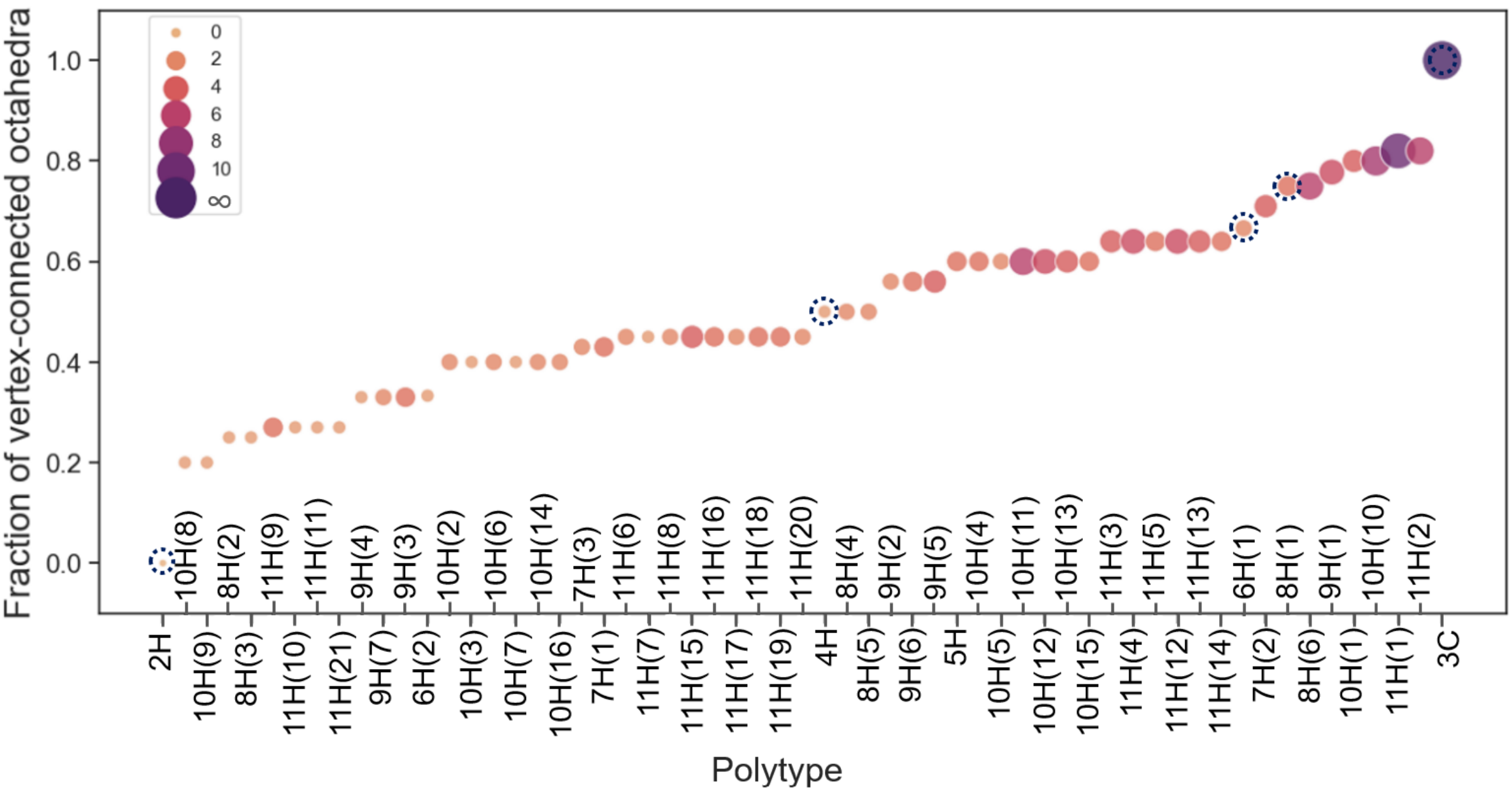

Figure 4. The intermediate close-packed polytypes between cubic perovskite (3C) and delta phase (2H) of $APbX_3$ with 4-11 close-packed layers with different fraction of vertex-connected $PbI_6$ octahedra. Dashed circles show experimentally known structures of polytypes. The size and color of the circles indicate the thickness of the "cubic block" in the structure.

The dependence of the calculated band gap on the number of layers of close packings with vertex-connected octahedra (*c* layers) is also non-linear (Figure 5). The thicker the block with cubic packing (vertex-connected octahedra) in the structure, the smaller the band gap. The larger sizes of the circle correspond to thicker «cubic blocks» in the close packing. For example, for 9-layer polytypes 9H(2) – |hch**cc**hhcc|, 9H(6) –|**ccc**hhchch|, and 9H(5) –|**cccc**hchhh| the proportion of layers with vertex-connected octahedra 9 (c) are the same, but the thickness of the «cubic block» in close packings increases from 2 to 3 and 4 respectively, which leads to a decrease in the values of the calculated band gap and, therefore, leads to a nonlinear trend in the dependence of the band gap on the number of vertex-connected octahedra in polytypes. Thus, there are clear dependence of the band gap on the number of vertex-connected octahedra in crystallographically possible close-packed $APbI_3$ polytypes (Figure 6) that are distinguished depending on the thickness of the «cubic block». It should be noted that the calculated band gap values are consistent either with the measured optical band gaps for a number of similar hexagonal polytypes containing tin as a metal ions [5] or with the experimentally known polytypes $FAPbX_3$ ($X=I^-$, $Br^-$) [3][2], in which the color of the crystals correlates with the band gap.

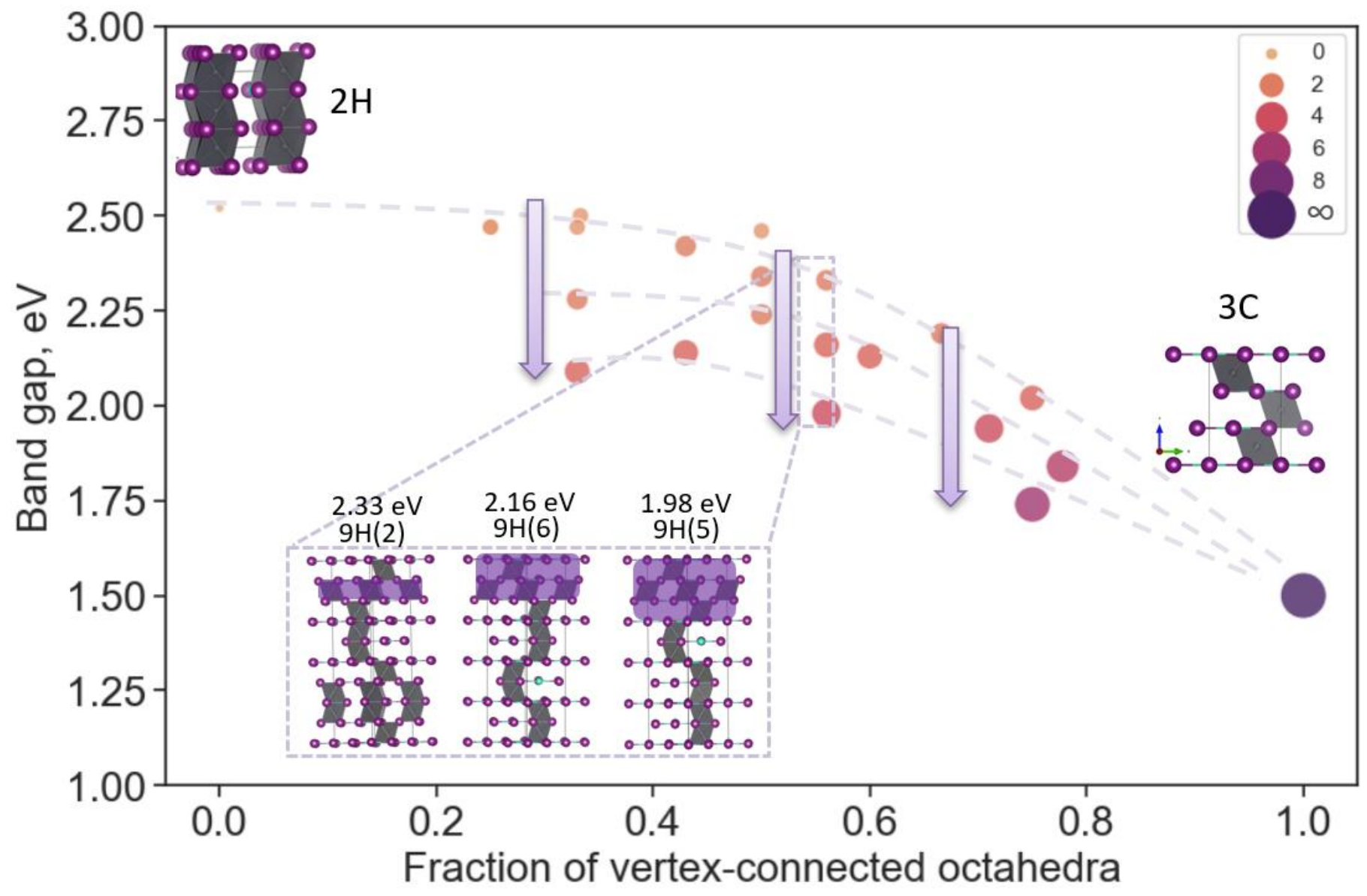

Figure 5. The band gaps and proportions of vertex-connected $PbI_6$ octahedra in hexagonal polytypes with 2-9 close-packed layers. The size of the balls shows the size of the «cubic» block of vertex-connected octahedra in the crystal structure of polytype.

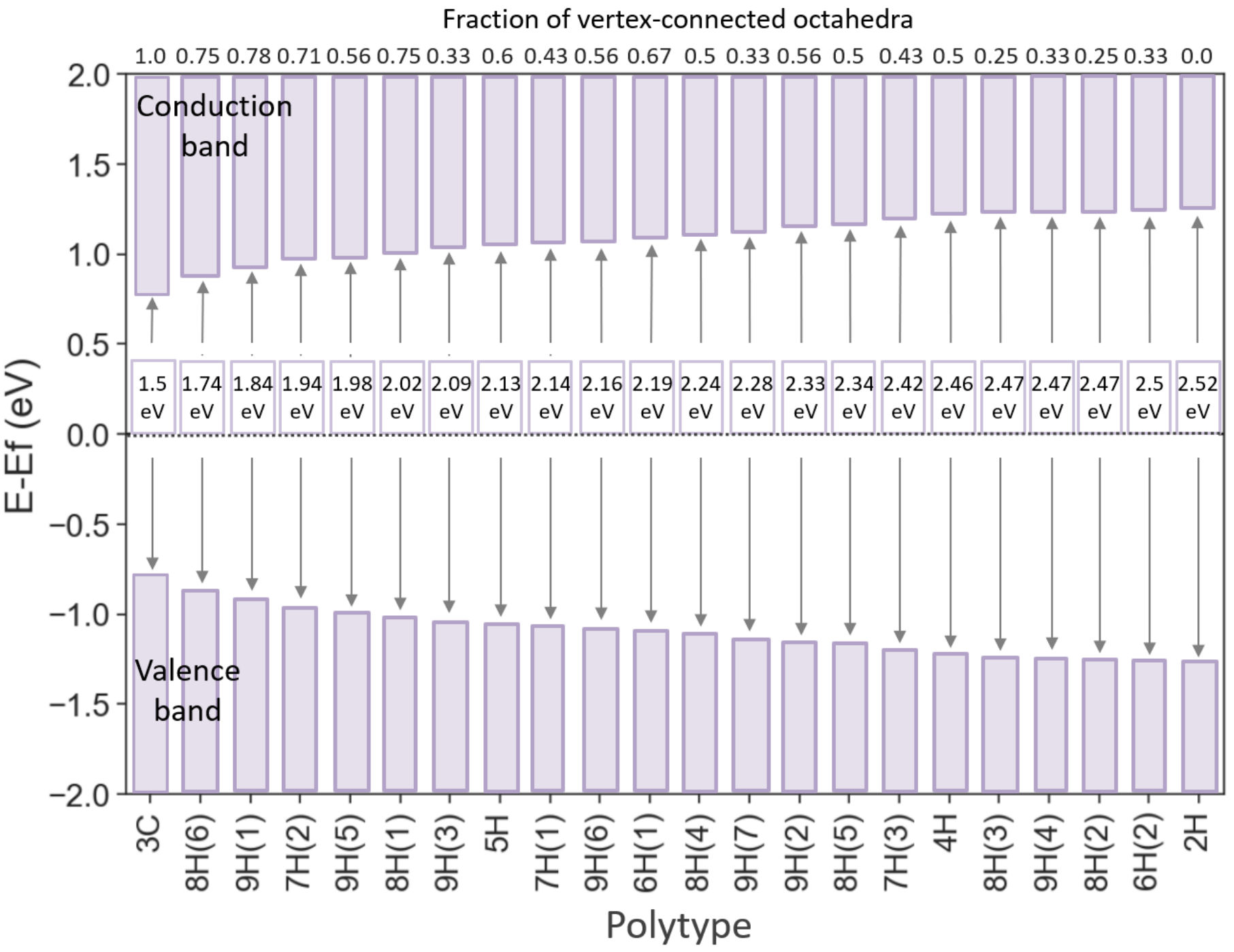

Figure 6. Energy level scheme for $APbI_3$ polytypes from DFT calculations without SOC.

The revealed dependences of the decrease in the band gap for various polytypes with an increase in the fraction of cubic layers and an increase in the thickness of the "cubic block" make it possible to predict either the electronic properties of higher polytypes (with n > 11) or defective materials with a broken sequence of layers [8][29][30]. Indeed, a microstructure of halide perovskite usually contains plethora of planar defects such as stacking faults (local deviation from the close-packed stacking sequences in the structure) or intergrowth structures. Considering the sequence of layers perpendicular to the defective plane, we also can describe it by the ratio of c and h layers and the thickness of "cubic blocks". This allows us to make some conclusions about the influence of such defects on the optoelectronic properties of the halide perovskites.

From the structural point of view, the density of the hexagonal stacking faults in cubic perovskite is described by the ratio of c and h layers. Analogously, the intergrowths with some hexagonal polymorph (δ-$FAPbI_3$) can be represented by thickness of hexagonal "blocks". According to the trends seen in Figures 5 and 6, the higher concentration of both the types of defects should lead to an increase of a band gap value locally. Therefore, in comparison with classical semiconductors [31], [32], planar defects in metal halide perovskites introduce no in-gap electronic states causing non-radiative recombination. Moreover, a local insertion of a wider bang gap phase may even reduce the non-radiative recombination by repelling charge carriers form such a defective boundary. Accordingly, such planar defects should have no detrimental effect on the optical properties of the material (intensity and lifetime of PL), but may even improve them. We also suggest that stacking-sequence-related planar defects should act as a typical heterostructures of bulk semiconductor with quantum wells (e.g. $GaAs/Al_xGa_{1-x}As$)[33]. Particularly, such defects are expected to be hardly detectable by optical methods at room temperature but they should manifest themselves at low temperatures as a weak blue-shifted optical bands in PL and absorption spectra, which could serve as a reasonable explanation of recently reported "intrinsic quantum confinement" in $FAPbI_3$ and $FA_{1-x}Cs_xPbI_3$ perovskites [34][35].

On the other hand, repelling charge carriers from the planar defects should impair the charge transport. In addition, as seen from the band diagrams for considered 9H polytypes (9H(2), 9H(6) and 9H(5)) (Figure S3), the reduction of the «cubic block» thickness results in lower dispersion near the VBM and CBM (Figure S3). Quantum confinement effect of individual h layers or polytypes in the perovskite (3C) matrix and reduced mobility should have negative consequences for the efficiency of devices. Therefore, we believe that fine control over the formation of different polytypes, their intergrowths and related stacking faults should be one of the main roads for further progress in efficiency of PSCs.

**Conclusions**

Using the group theory approach, we revealed all possible crystal structures of 59 $APbI_3$ iodoplumbate polytypes with a number of close-packed layers up to 11. All the 54 predicted structures and several experimentally observed ones (3C, 2H, 4H, 6H, 8H) were linked by subgroup-supergroup symmetry relationships and are grouped according with the main topological parameters such as the ratio of cubic (*c*) and hexagonal (*h*) close-packed layers and the thickness of blocks of cubic layers in the structures. Using DFT calculations, we revealed the most important relationships between crystal and electronic structures of the polytypes. We showed an increasing scatter of band gap values with an increase of the number of close-packed layers in polytypes. Two main fundamental factors affecting the band gap ($E_g$) were found to be the ratio of cubic and hexagonal layers (r) setting the general trend of the $E_g$ decrease with r and the thickness of blocks of adjacent cubic layers in the structures introducing a quantum confinement effect. For each the given thickness of cubic block in the structure of polytype, the monotonic decrease of band gap with increase of r was observed. We also generalized the obtained dependences to the case of broken sequence of close-packed layers which is observed in real materials containing planar defects such as stacking faults or intergrowth. The local structure of such defects implies that they should have the same electronic properties as the bulk polytypes with the same sequence of h and c layers. We believe that such planar defects have no detrimental effect on the bulk optical properties of the material (intensity and lifetime of PL) but, oppositely, may improve desired optical characteristics by repelling charge carriers form grain boundaries and inhibiting their recombination. However, the presence of such polytypes in the perovskite (3C) matrix could lead to a decrease in the charge carrier mobility and obstacle the charge transport, which might have negative consequences for the efficiency of solar cell devices.

**AUTHOR INFORMATION**

Corresponding Author * alexey.bor.tarasov@yandex.ru

**Author Contributions**

The manuscript was written through contributions of all authors. All authors have given approval to the final version of the manuscript.

**Notes**

The authors declare no competing financial interest.

**ACKNOWLEDGMENT**

This work was financial supported by a grant from the Russian Science Foundation, project number 19-73-30022.

**Structure-related bandgap of hybrid lead halide perovskites and close-packed $APbX_3$ family of phases**

Ekaterina I. Marchenko [a,c], Sergey A. Fateev [a], Vadim V.Korolev [b], Vladimir Buchinskii [c], Eremin N.N.[c], Eugene A. Goodilin[a,b] and Alexey B. Tarasov*[a,b]

[a] *Laboratory of New Materials for Solar Energetics, Department of Materials Science, Lomonosov Moscow State University; 1 Lenin Hills, 119991, Moscow, Russia;*
*e-mail:alexey.bor.tarasov@yandex.ru*
[b] *Department of Chemistry, Lomonosov Moscow State University; 1 Lenin Hills, 119991, Moscow, Russia*
[c] *Department of Geology, Lomonosov Moscow State University; 1 Lenin Hills, 119991, Moscow, Russia*
[*] corresponding author

# Supporting information

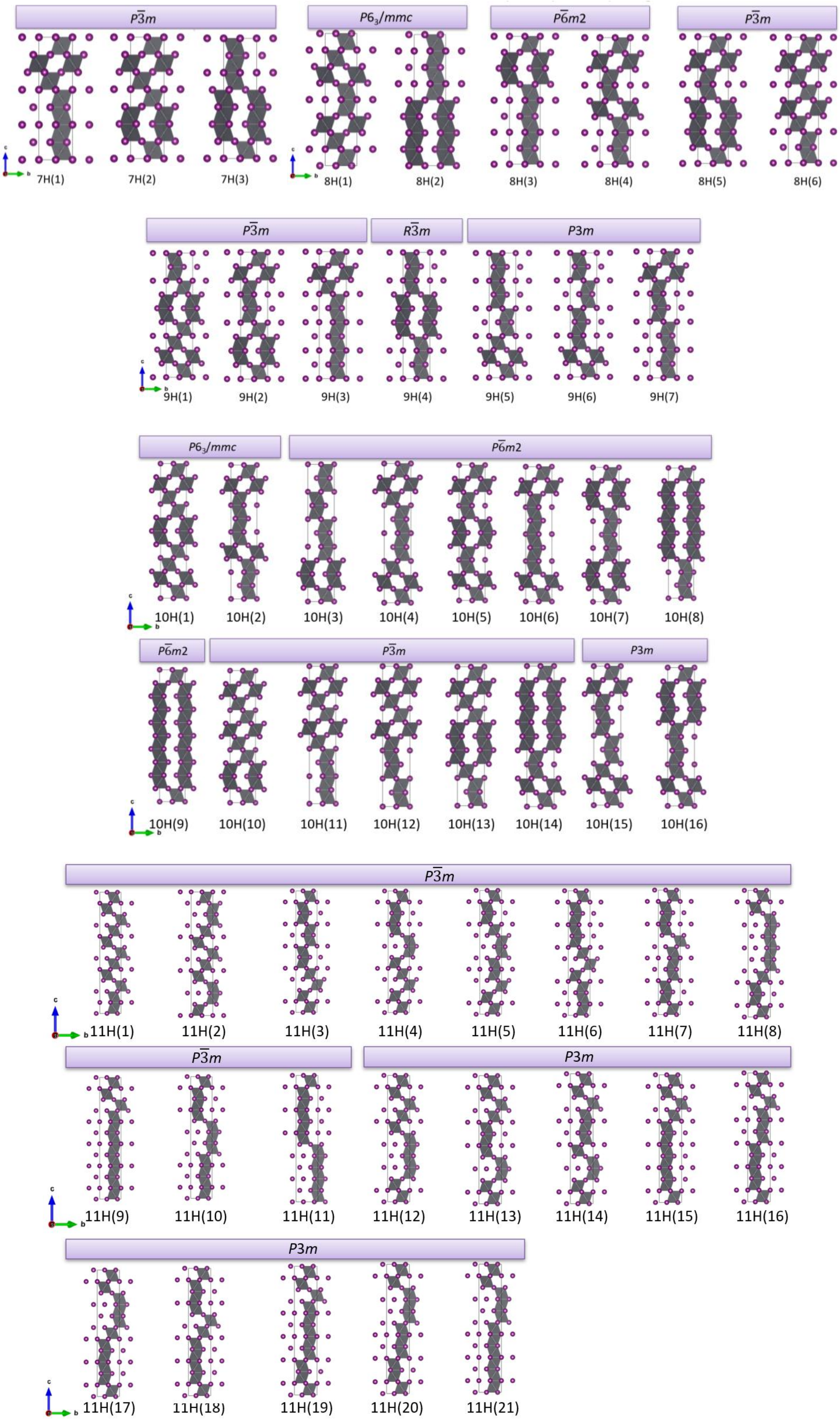

Figure S1. Constructed all possible 7H, 8H 9H, 10H and 11H close-packed polytypes of $APbI_3$.

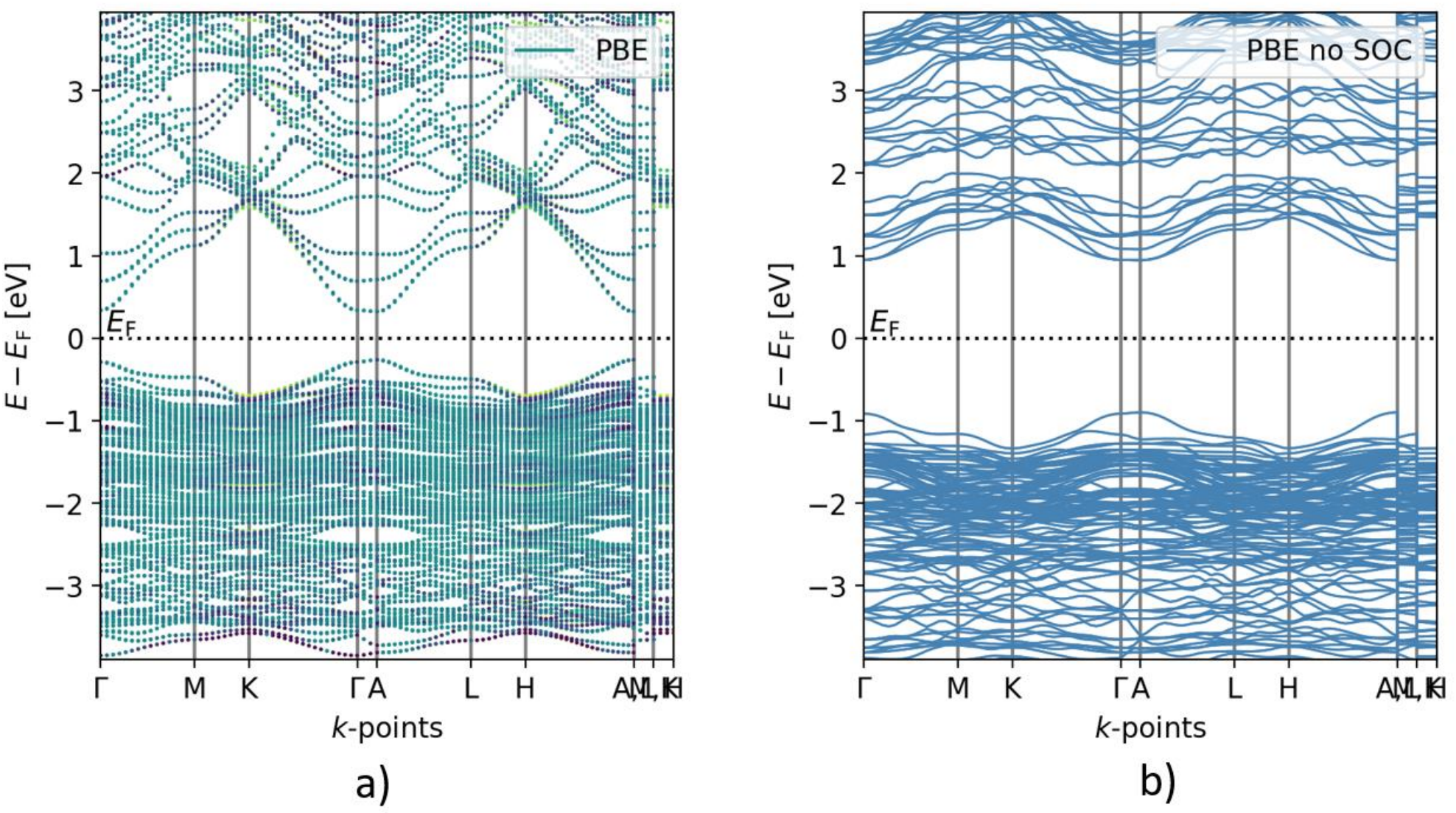

Figure S2. Band structure of the 9H(1) polytype computed at the PBE+SOC (a) and PBE (b) levels of theory.

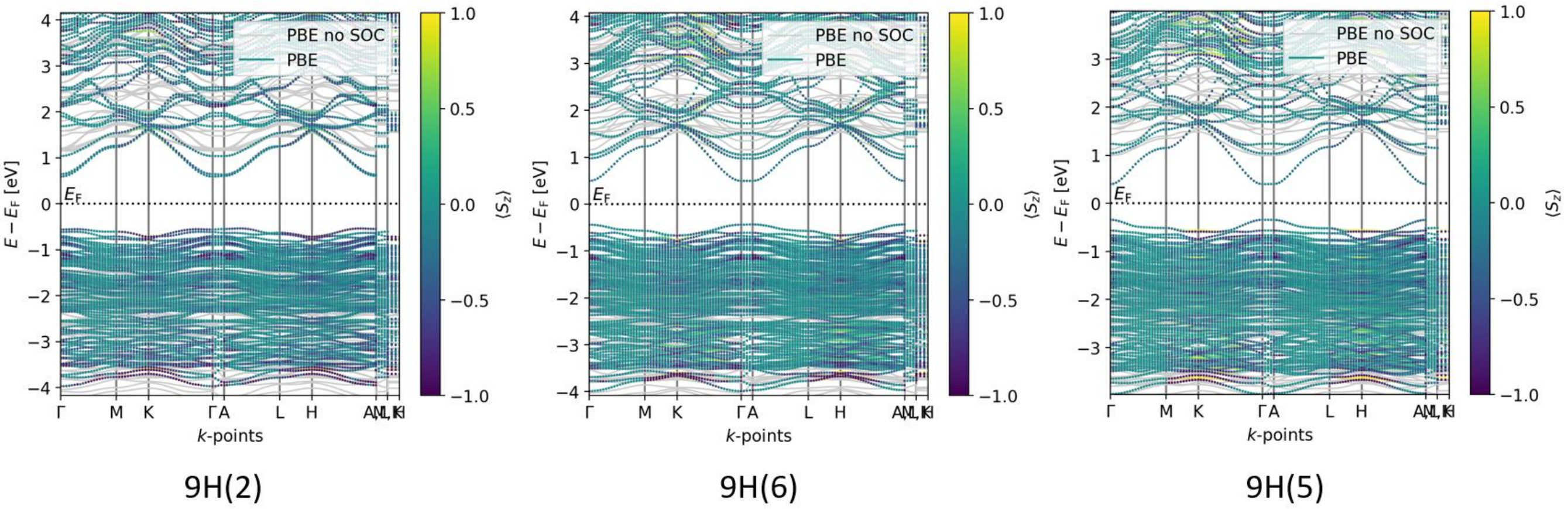

Figure S3. The band structures of 9H polytypes with different proportion of *h* and *c* close-packed layers. Gray color shows calculations without SOC, color shows results of calculations with SOC.

Table S1. Characteristics of possible configurations of close packings from 2 to 9 layers for $APbI_3$.

| A number of layers in close packing | ABC symbol | Space group symmetry of the close packing | Space group symmetry of structures occupied octahedral voids by lead ions | Fraction of closed-packed layers where $PbI_6$ octahedra connected by vertices |
|---|---|---|---|---|
| 10H(1) | (ABCABCBACB) | \|hccchcccc\| | $D_{6h}^4$ ($P6_3/mmc$) | 0.8 |
| 10H(2) | (ABABCACACB) | \|hhhcchhhcc\| | $D_{6h}^4$ ($P6_3/mmc$) | 0.4 |
| 10H(3) | (ABCBABACAB) | \|hchchhchch\| | $D_{3h}^1$($P\overline{6}m2$) | 0.4 |
| 10H(4) | (ABCABABACB) | \|hccchhhccc\| | $D_{3h}^1$($P\overline{6}m2$) | 0.6 |
| 10H(5) | (ABCACBCACB) | \|hcchchchcc\| | $D_{3h}^1$($P\overline{6}m2$) | 0.6 |
| 10H(6) | (ABCACACACB) | \|hcchhhhhcc\| | $D_{3h}^1$($P\overline{6}m2$) | 0.4 |
| 10H(7) | (ABCBABABCB) | \|hchchhhchc\| | $D_{3h}^1$($P\overline{6}m2$) | 0.4 |
| 10H(8) | (ABABCBCBCB) | \|hhhchhhhhc\| | $D_{3h}^1$($P\overline{6}m2$) | 0.2 |
| 10H(9) | (ABCBCBCBCB) | \|hchhhhhhhc\| | $D_{3h}^1$($P\overline{6}m2$) | 0.2 |
| 10H(10) | (ABCBACBACB) | \|hchcccccc\| | $D_{3d}^3$ ($P\overline{3}m$) | 0.8 |
| 10H(11) | (ABABACBACB) | \|hhhhcccccc\| | $D_{3d}^3$ ($P\overline{3}m$) | 0.6 |
| 10H(12) | (ABACACBACB) | \|hhchhccccc\| | $D_{3d}^3$ ($P\overline{3}m$) | 0.6 |
| 10H(13) | (ABACBCBACB) | \|hhcchhcccc\| | $D_{3d}^3$ ($P\overline{3}m$) | 0.6 |
| 10H(14) | (ABCACBCBCB) | \|hcchchhhhc\| | $D_{3d}^3$ ($P\overline{3}m$) | 0.4 |
| 10H(15) | (ABCABACACB) | \|hccchchhcc\| | $C_{3v}^1$ ($P3m$) | 0.6 |
| 10H(16) | (ABCACACBCB) | \|hcchhhchhc\| | $C_{3v}^1$ ($P3m$) | 0.4 |
| 11H(1) | (ABCBACBACBC) | \|hcccccccccch\| | $D_{3d}^3$ ($P\overline{3}m$) | 0.82 |
| 11H(2) | (ACBCABCABCB) | \|cchcccccchc\| | $D_{3d}^3$ ($P\overline{3}m$) | 0.82 |
| 11H(3) | (ACBCACBCACB) | \|cchchchchcc\| | $D_{3d}^3$ ($P\overline{3}m$) | 0.64 |
| 11H(4) | (ABCBACBACAB) | \|hchcccccchch\| | $D_{3d}^3$ ($P\overline{3}m$) | 0.64 |
| 11H(5) | (ACABACBABCB) | \|chchccchchc\| | $D_{3d}^3$ ($P\overline{3}m$) | 0.64 |
| 11H(6) | (ACBCBCACACB) | \|cchhhchhhcc\| | $D_{3d}^3$ ($P\overline{3}m$) | 0.45 |
| 11H(7) | (ACACBCACBCB) | \|chhchchchhc\| | $D_{3d}^3$ ($P\overline{3}m$) | 0.45 |
| 11H(8) | (ABACBCACBAB) | \|hhcchchcchh\| | $D_{3d}^3$ ($P\overline{3}m$) | 0.45 |
| 11H(9) | (ABABACBABAB) | \|hhhhccchhhh\| | $D_{3d}^3$ ($P\overline{3}m$) | 0.27 |
| 11H(10) | (ABACACBCBAB) | \|hhchhchhchh\| | $D_{3d}^3$ ($P\overline{3}m$) | 0.27 |
| 11H(11) | (ACACACBCBCB) | \|chhhhchhhhc\| | $D_{3d}^3$ ($P\overline{3}m$) | 0.27 |
| 11H(12) | (ABCACACBACB) | \|hcchhhccccc\| | $C_{3v}^1$ ($P3m$) | 0.64 |
| 11H(13) | (ABCACBCBACB) | \|hcchchhcccc\| | $C_{3v}^1$ ($P3m$) | 0.64 |
| 11H(14) | (ABCACBABACB) | \|hcchcchhccc\| | $C_{3v}^1$ ($P3m$) | 0.64 |
| 11H(15) | (ABCBCBCBACB) | \|hchhhhhcccc\| | $C_{3v}^1$ ($P3m$) | 0.45 |
| 11H(16) | (ABCBCBABACB) | \|hchhhchhccc\| | $C_{3v}^1$ ($P3m$) | 0.45 |
| 11H(17) | (ABCBCBACACB) | \|hchhhcchhcc\| | $C_{3v}^1$ ($P3m$) | 0.45 |
| 11H(18) | (ABCBCBACBCB) | \|hchhhccchhc\| | $C_{3v}^1$ ($P3m$) | 0.45 |
| 11H(19) | (ABCBABABACB) | \|hchchhhhccc\| | $C_{3v}^1$ ($P3m$) | 0.45 |
| 11H(20) | (ABCBABACACB) | \|hchchhchhcc\| | $C_{3v}^1$ ($P3m$) | 0.45 |
| 11H(21) | (ACACACBCBCB) | \|chhhhchhhhc\| | $C_{3v}^1$ ($P3m$) | 0.27 |

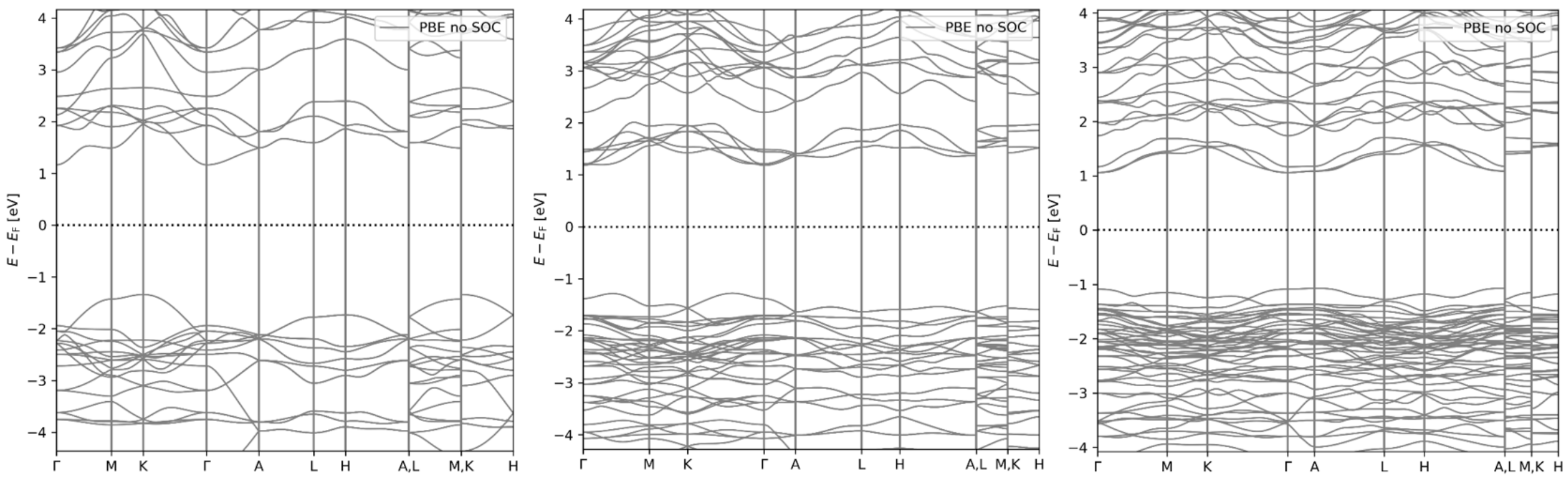

| | 2H | 4H | 5H |
|---|---|---|---|
| Band gap, eV | 2.52 | 2.46 | 2.13 |
| Band gap (direct), eV | 2.95 | 2.51 | 2.15 |

Figure S4. Calculated band structure without SOC for possible 2H, 4H and 5H polytypes of $APbI_3$.

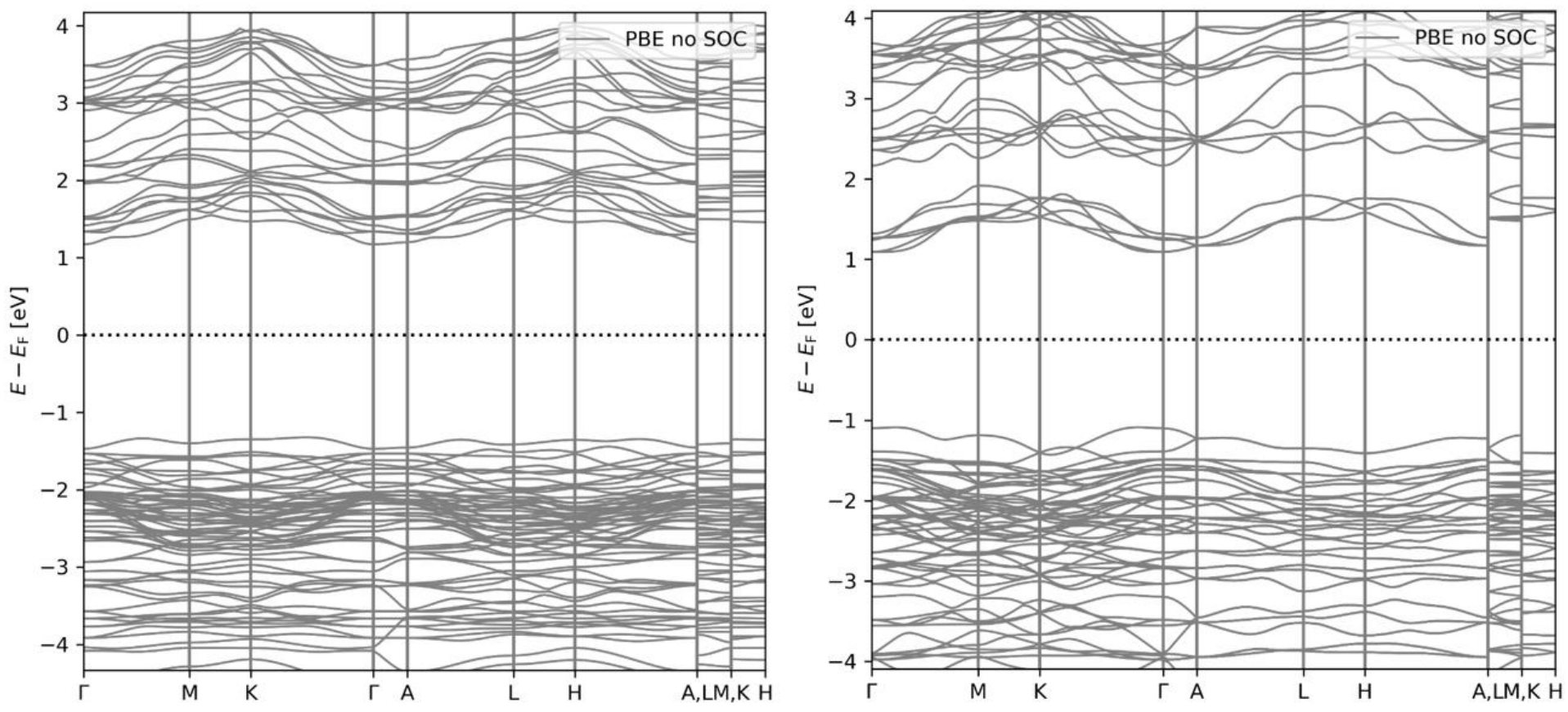

| | 6H(1) | 6H(2) |
|---|---|---|
| Band gap, eV | 2.5 | 2.19 |
| Band gap (direct), eV | 2.64 | 2.19 |

Figure S5. Calculated band structure without SOC for possible 6H polytypes of $APbI_3$.

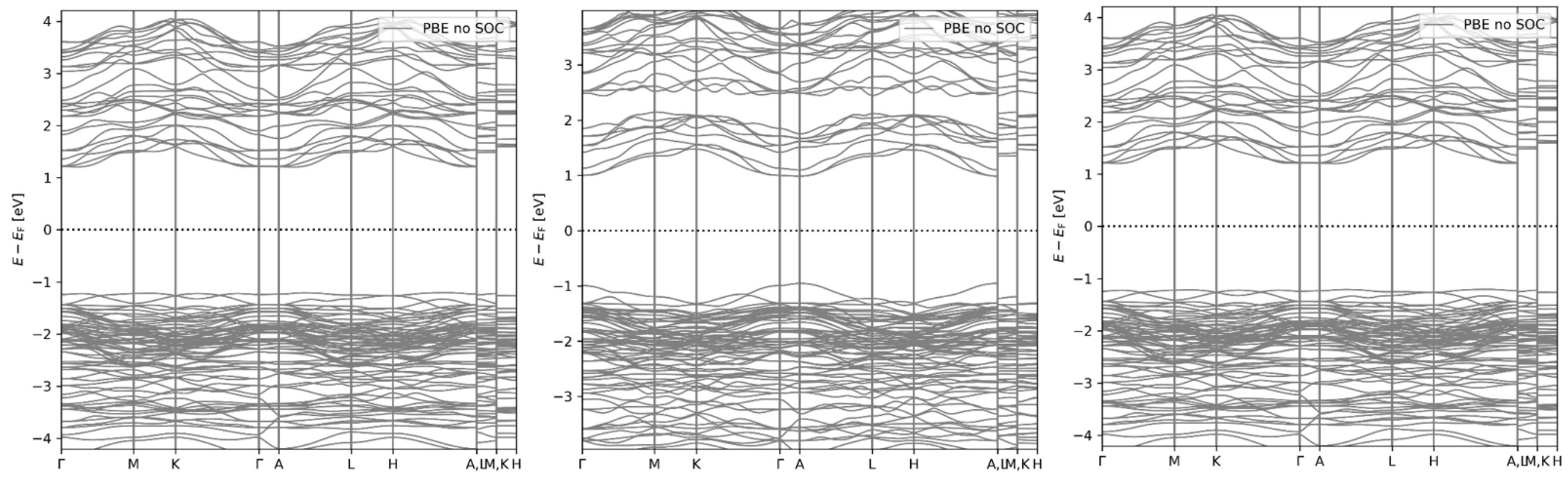

| | 7H(1) | 7H(2) | 7H(3) |
|---|---|---|---|
| Band gap, eV | 2.42 | 1.94 | 2.42 |
| Band gap (direct), eV | 2.42 | 1.94 | 2.42 |

Figure S6. Calculated band structure without SOC for possible 7H polytypes of $APbI_3$.

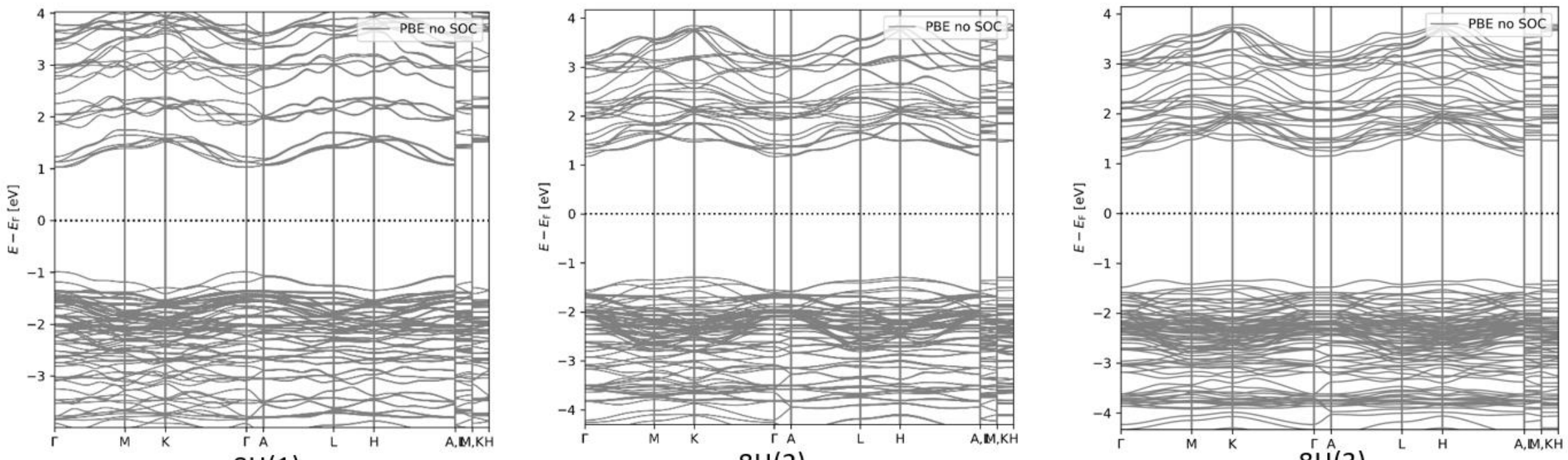

| | 8H(1) | 8H(2) | 8H(3) |
|---|---|---|---|
| Band gap, eV | 2.02 | 2.47 | 2.47 |
| Band gap (direct), eV | 2.02 | 2.70 | 2.62 |

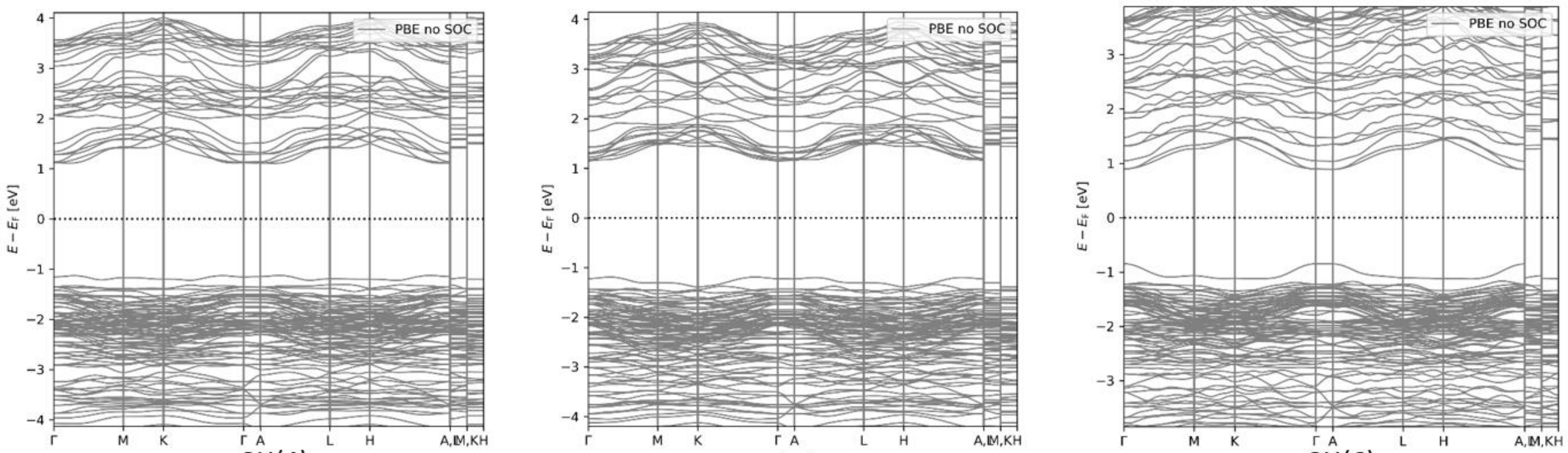

| | 8H(4) | 8H(5) | 8H(6) |
|---|---|---|---|
| Band gap, eV | 2.24 | 2.34 | 1.74 |
| Band gap (direct), eV | 2.24 | 2.37 | 1.74 |

Figure S7. Calculated band structure without SOC for possible 8H polytypes of $APbI_3$.

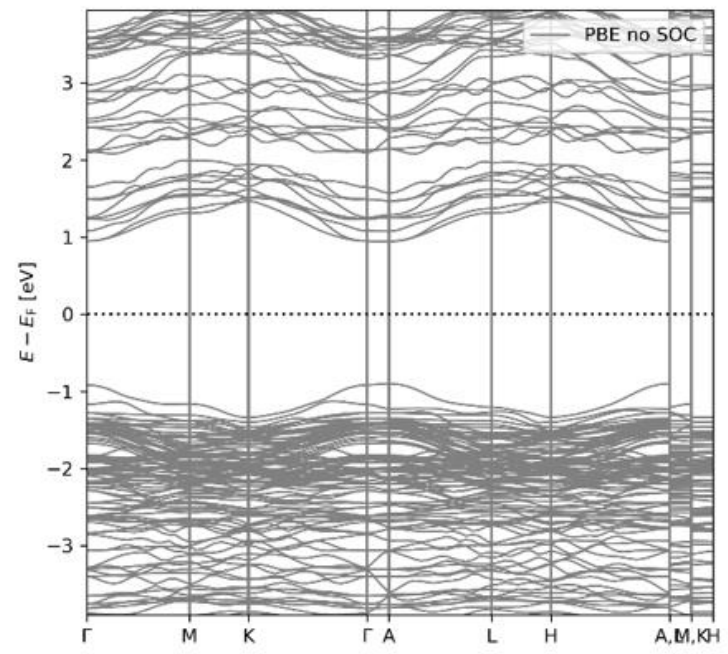

9H(1)

| | 9H(1) | 9H(2) | 9H(3) |
|---|---|---|---|
| Band gap, eV | 1.84 | 2.33 | 2.09 |
| Band gap (direct), eV | 1.84 | 2.34 | 2.15 |

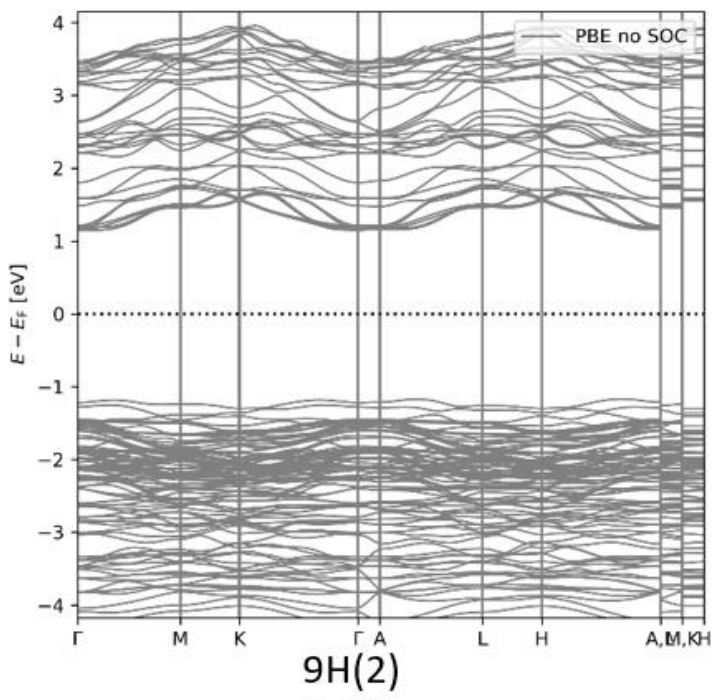

9H(2)

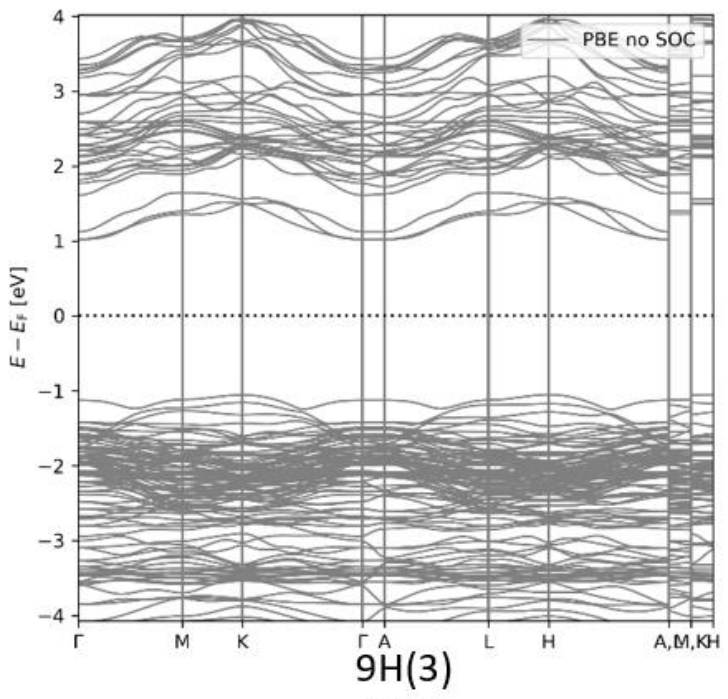

9H(3)

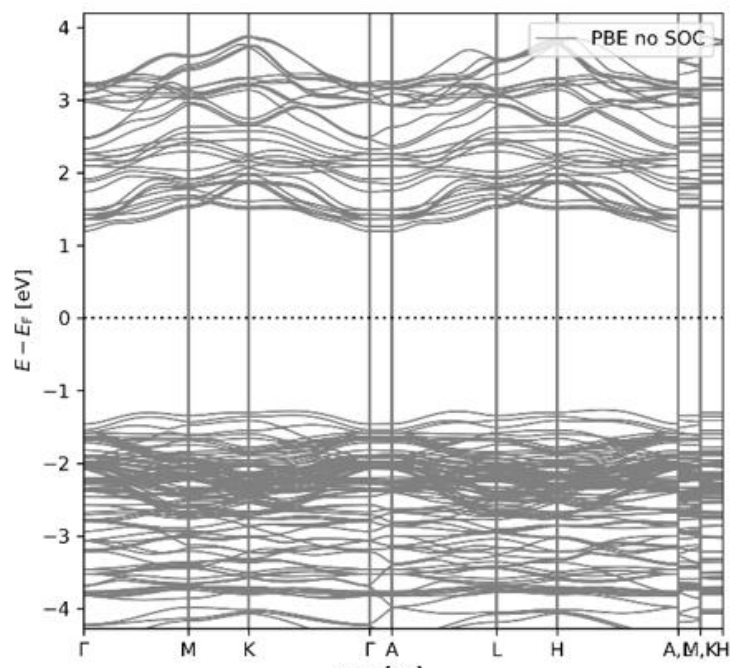

9H(4)

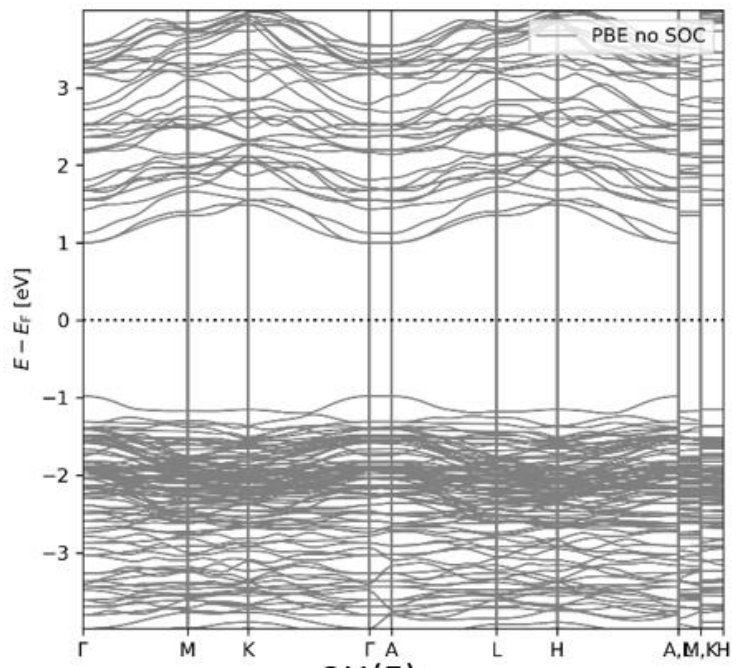

9H(5)

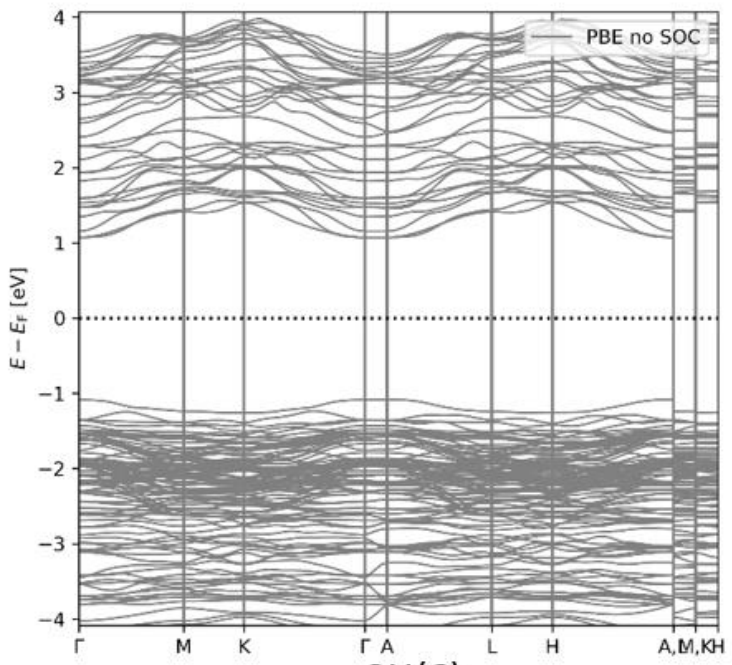

9H(6)

| | 9H(4) | 9H(5) | 9H(6) |
|---|---|---|---|
| Band gap, eV | 2.47 | 1.98 | 2.16 |
| Band gap (direct), eV | 2.62 | 1.98 | 2.16 |

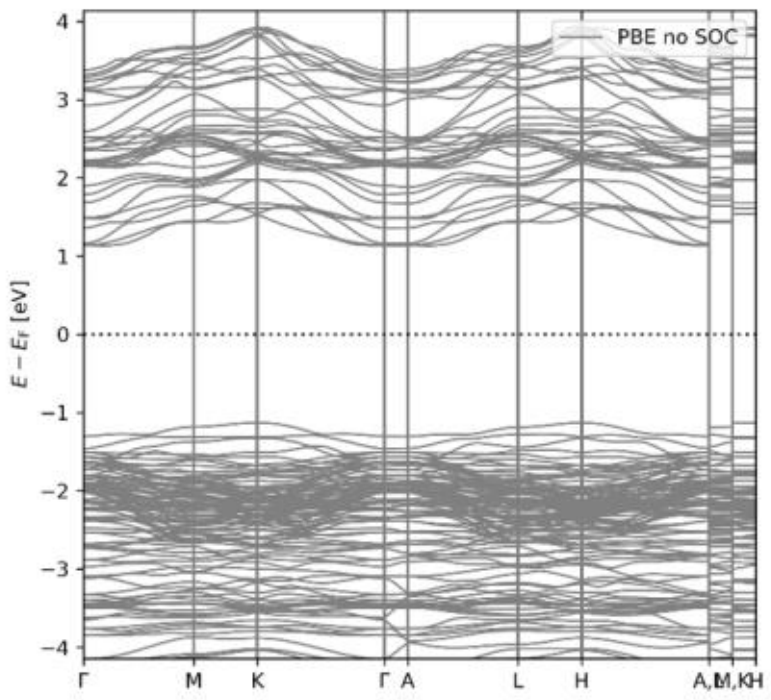

9H(7)

| | 9H(7) |
|---|---|
| Band gap, eV | 2.28 |
| Band gap (direct), eV | 2.28 |

Figure S8. Calculated band structure without SOC for possible 9H polytypes of $APbI_3$.

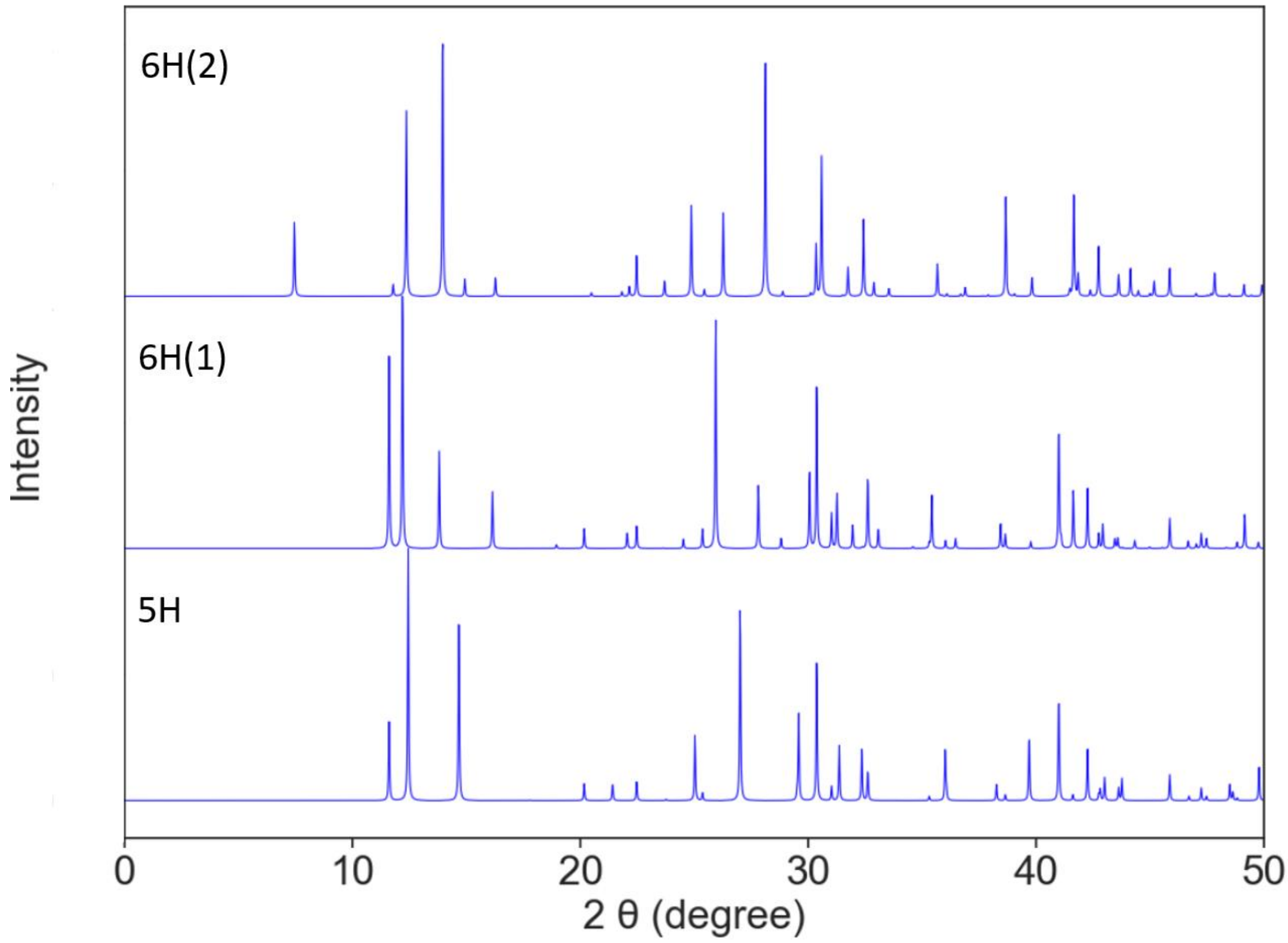

Figure S9. Calculated XRD pattern for 5H and 6H polytypes.

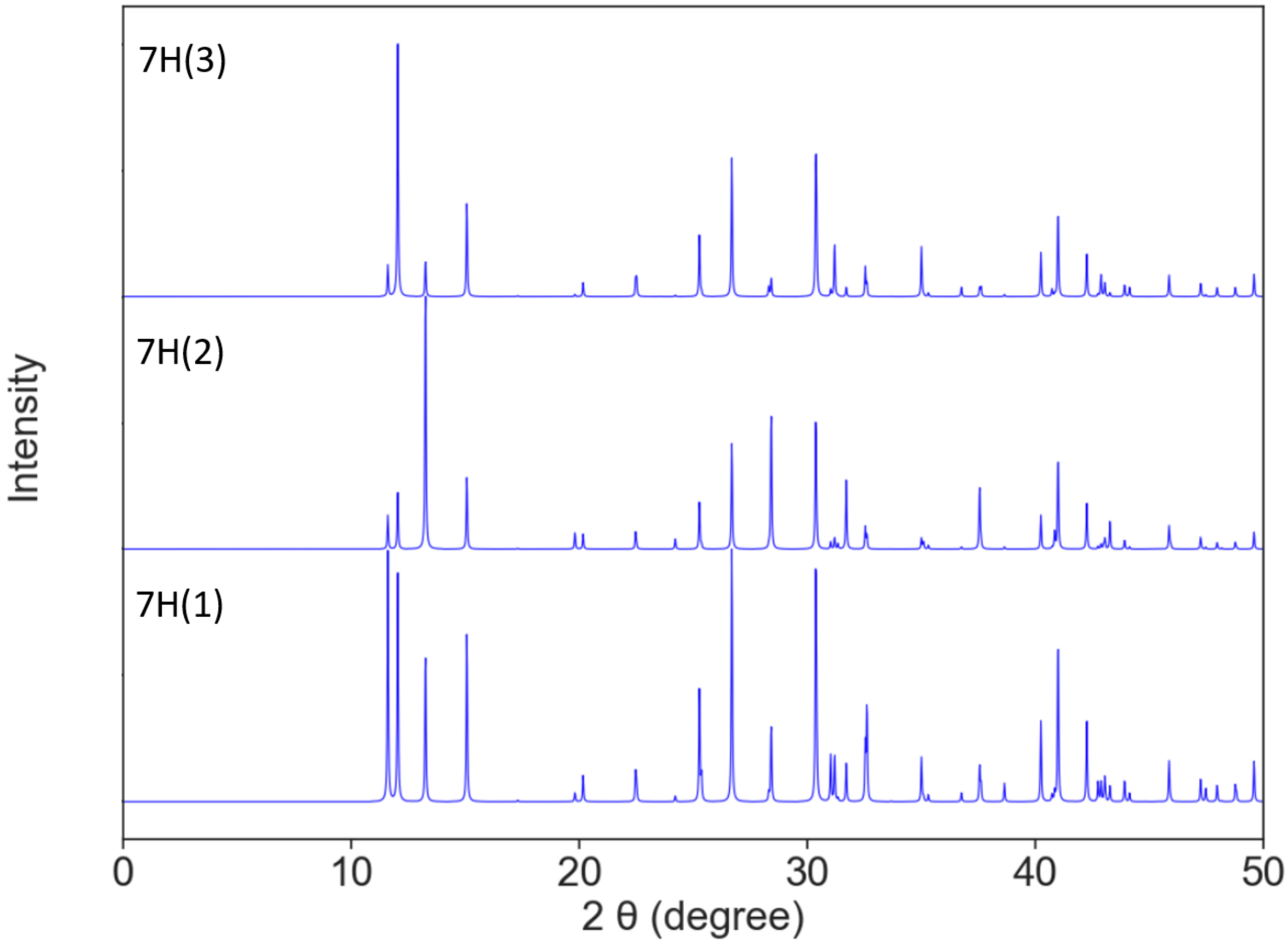

Figure S10. Calculated XRD pattern for 7H polytypes.

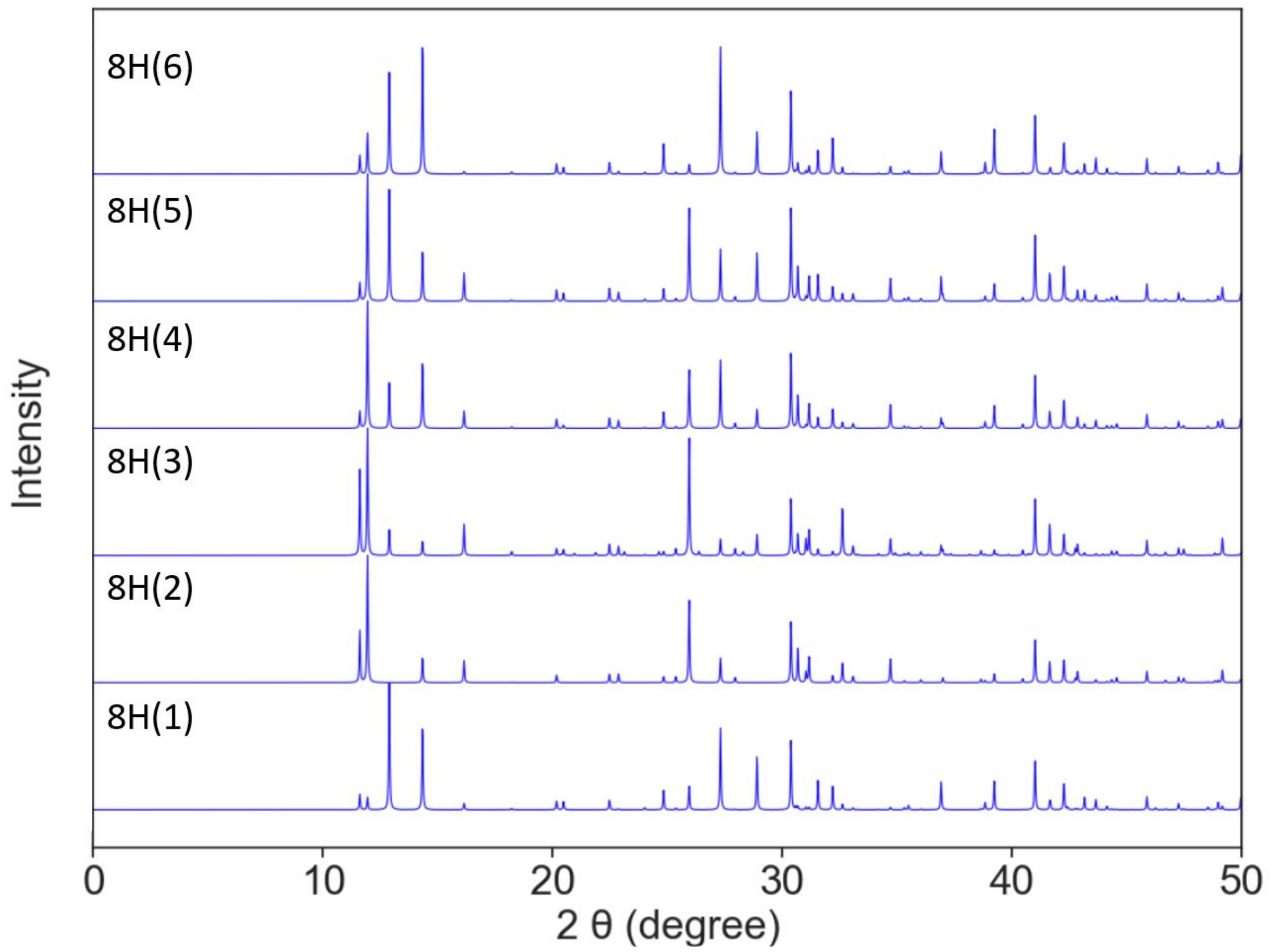

Figure S11. Calculated XRD pattern for 8H polytypes.

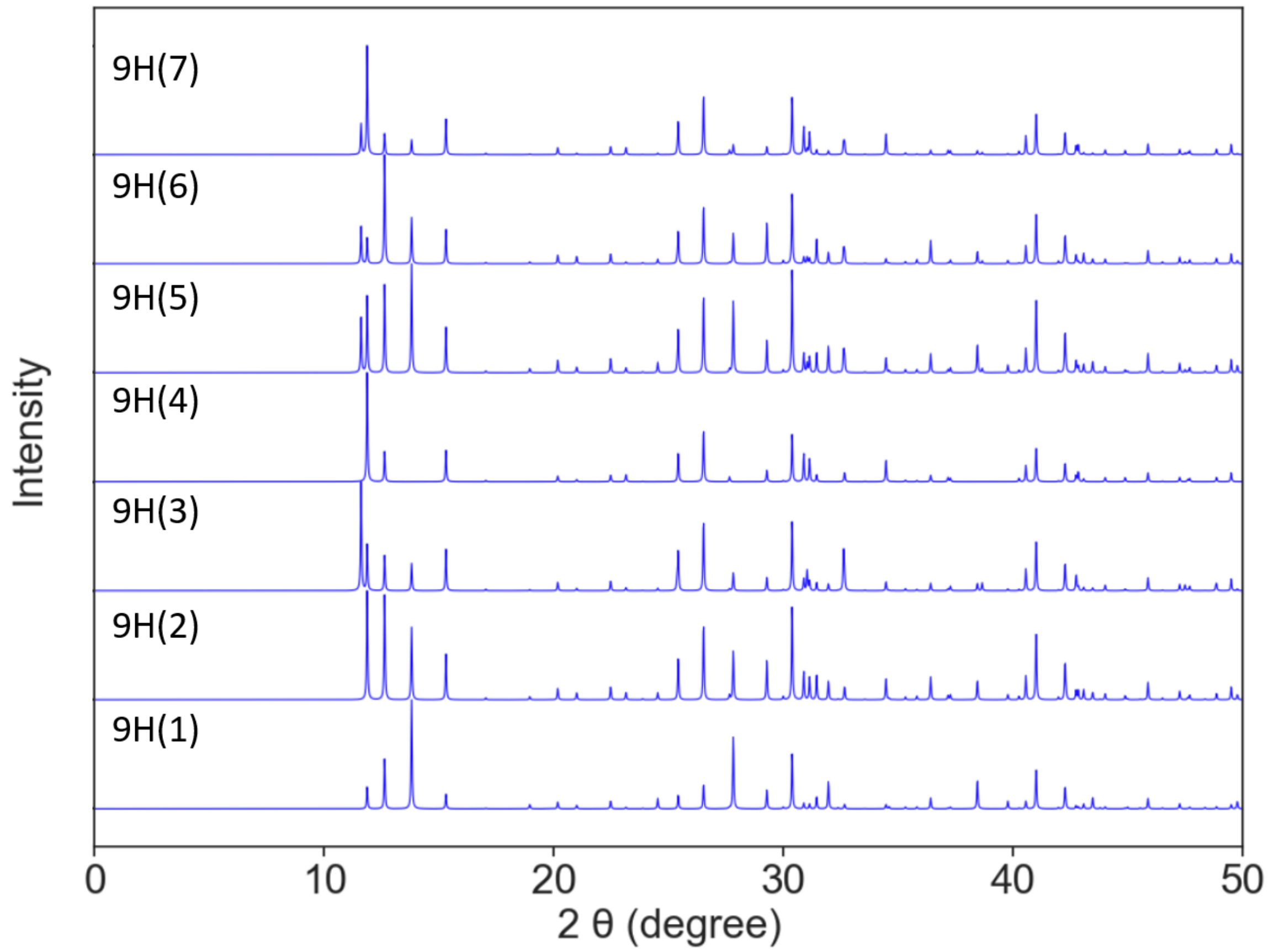

Figure S12. Calculated XRD pattern for 9H polytypes.

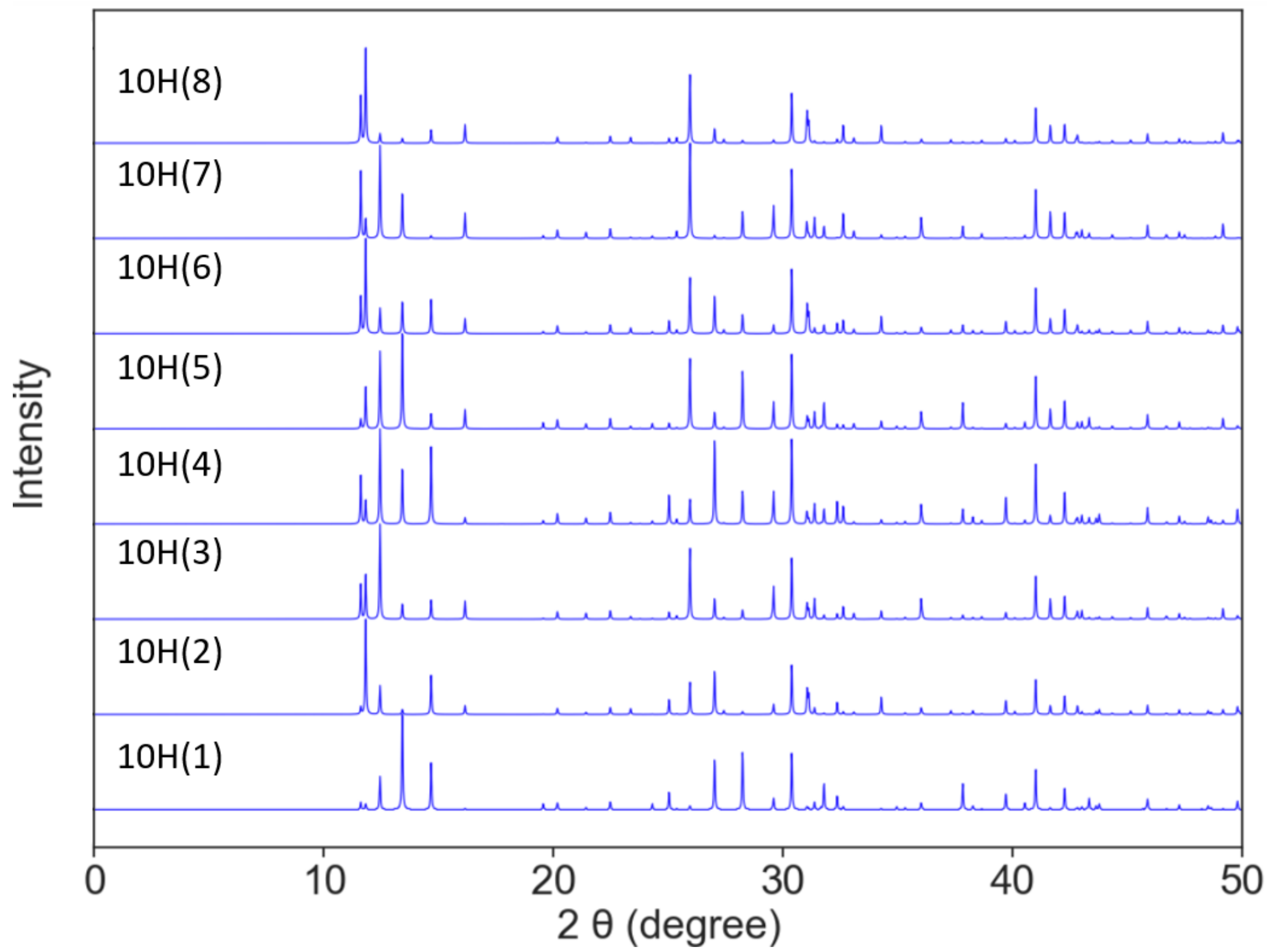

Figure S13a. Calculated XRD pattern for 10H (1-8) polytypes.

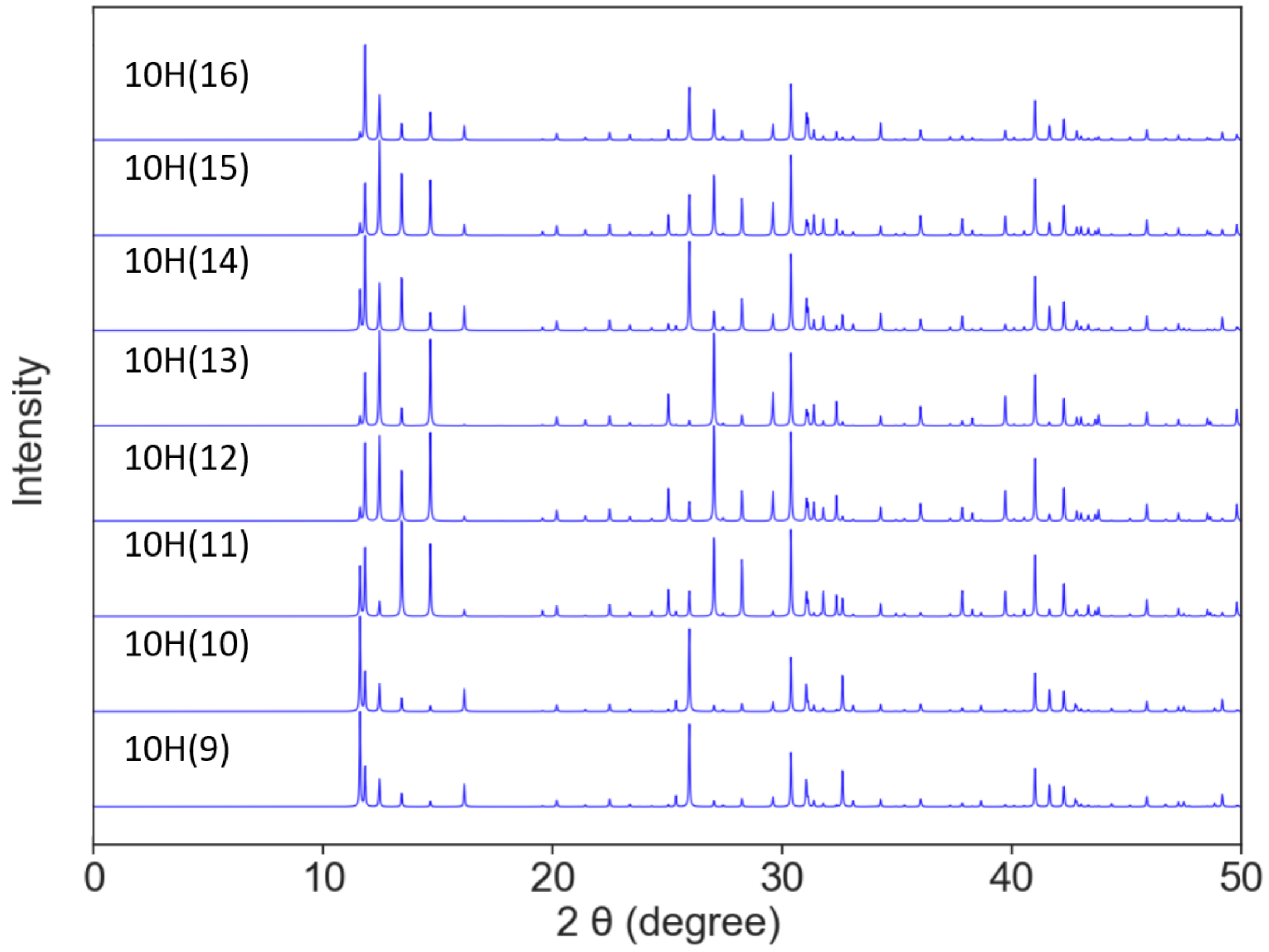

Figure S13b. Calculated XRD pattern for 10H (9-16) polytypes.

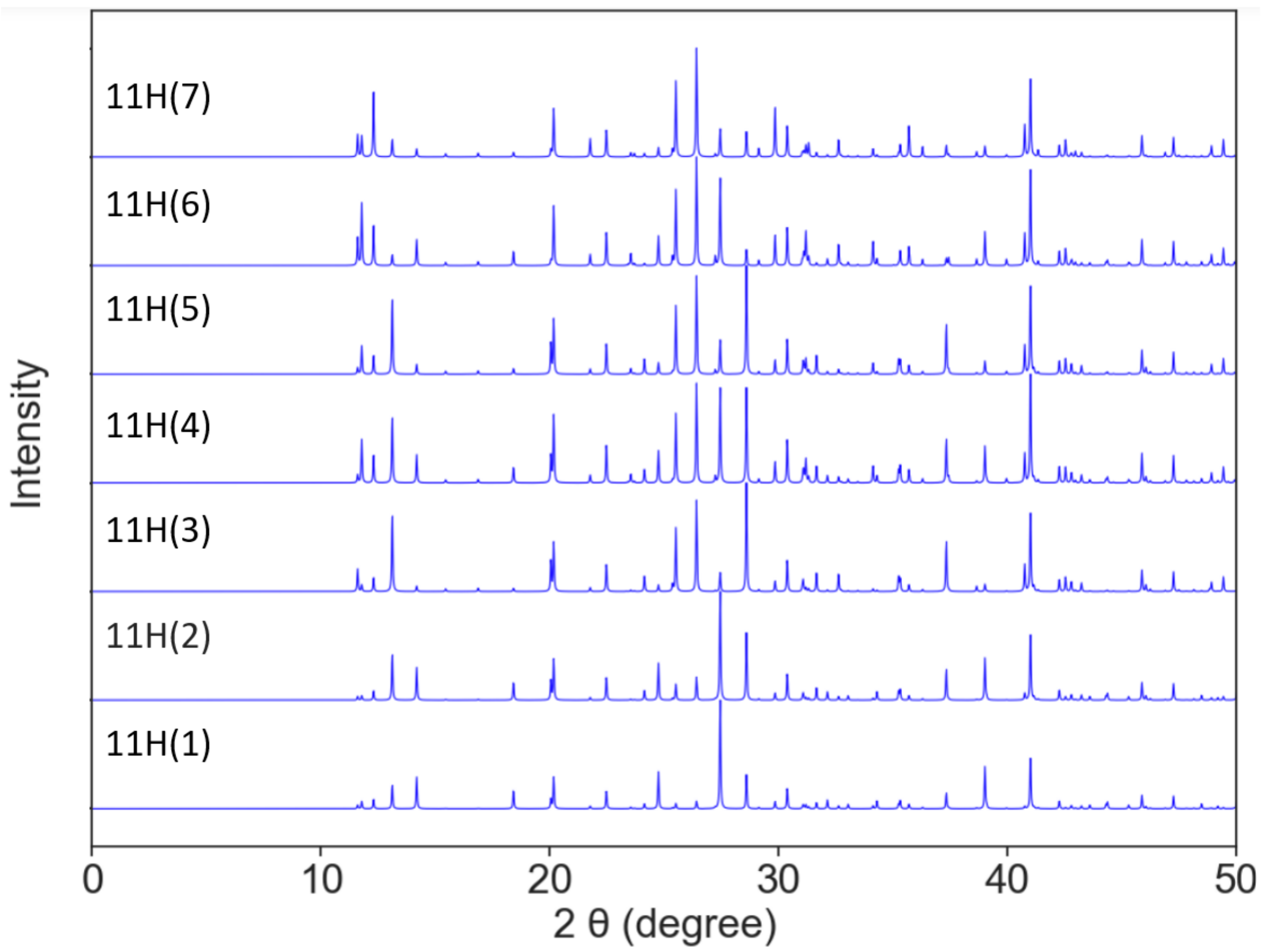

Figure S14a. Calculated XRD pattern for 11H (1-7) polytypes.

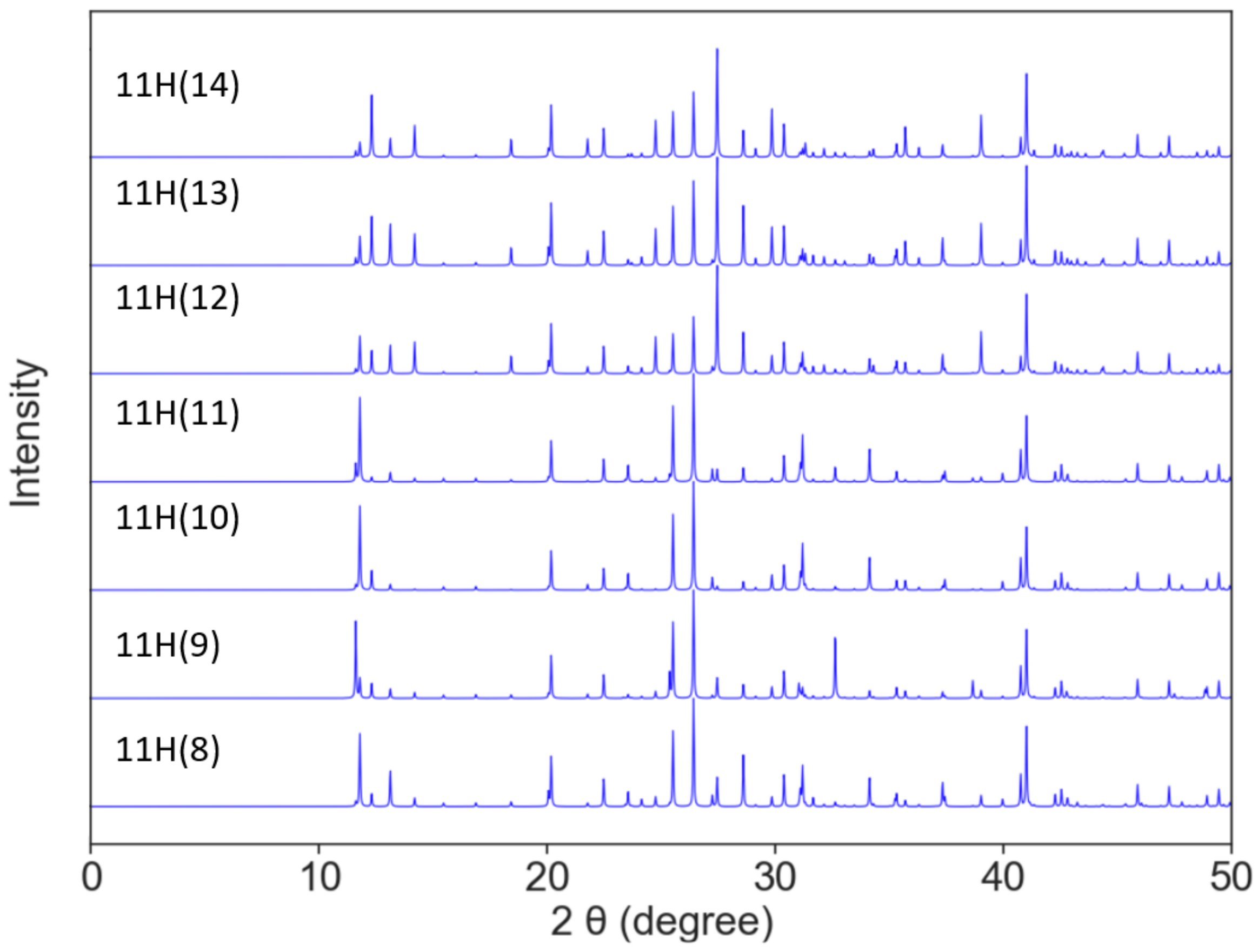

Figure S14b. Calculated XRD pattern for 11H (8-14) polytypes.

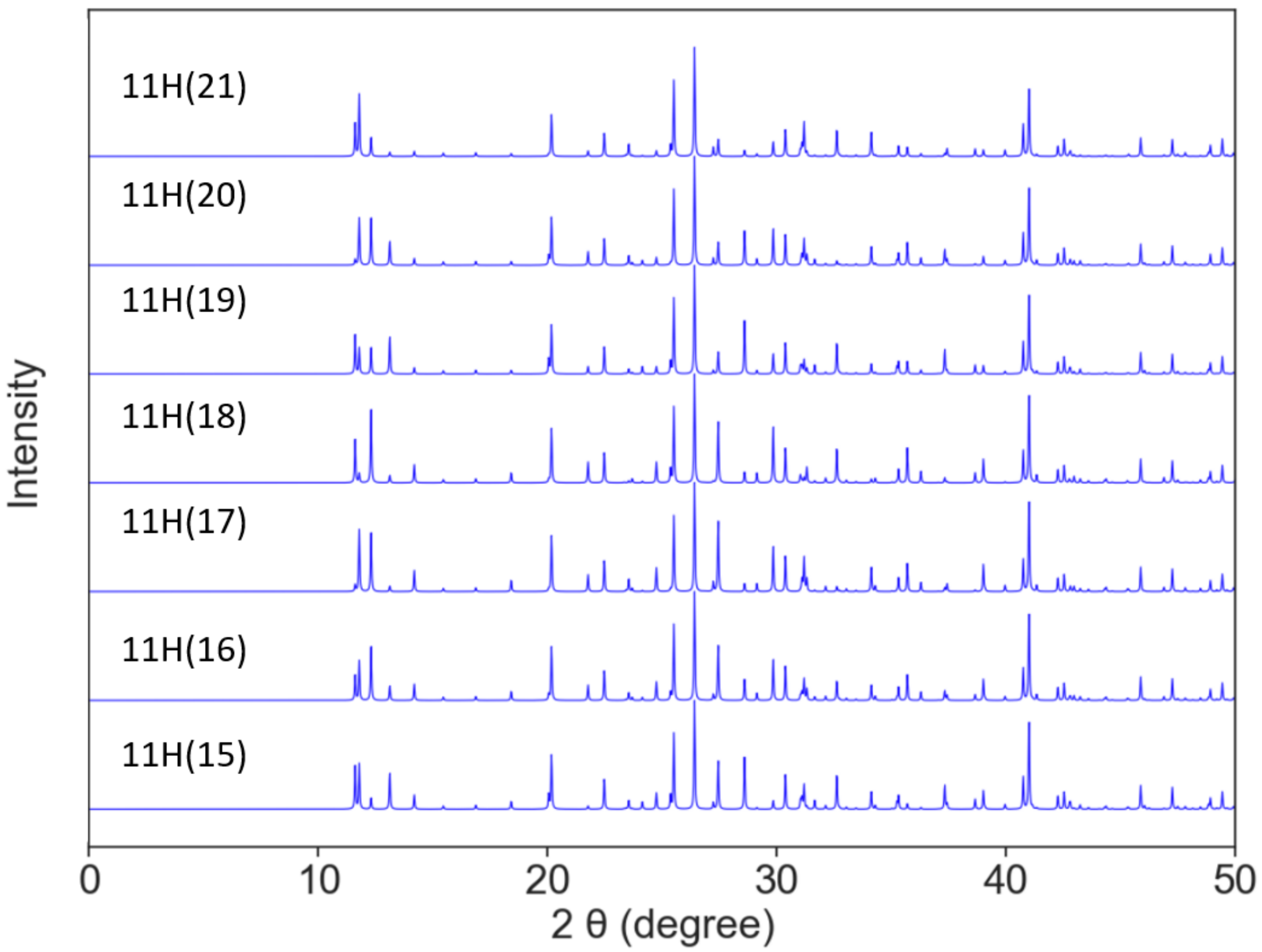

Figure S14c. Calculated XRD pattern for 11H (15-21) polytypes.